\documentclass[aps,floatfix,prd,nofootinbib,superscriptaddress,reprint,showpacs,10pt,preprintnumbers,longbibliography]{revtex4-2}
\usepackage[utf8]{inputenc}
\usepackage[pdftex]{graphicx}
\usepackage{float}
\usepackage{amsmath}
\usepackage{amssymb}
\usepackage{mathtools}
\usepackage{siunitx}
\usepackage[T1]{fontenc}
\usepackage{lmodern}
\usepackage{array}
\usepackage{bm}
\usepackage{mathrsfs}
\usepackage{dsfont} 
\usepackage{pifont}
\usepackage{multirow}
\usepackage{upgreek}
\usepackage{CJKutf8}
\usepackage{verbatim}
\usepackage{slashed}
\usepackage{ucs}
\usepackage{subfigure}
\usepackage{braket}
\usepackage{simplewick}
\usepackage[dvipsnames]{xcolor}
\usepackage{colortbl}
\usepackage{pgf}
\usepackage{tabularx}
\usepackage[
  pdftex,
  pdftitle={Lambda baryon excited states at SU3 point},
  pdfauthor={},
  bookmarks,
  colorlinks,
  linkcolor=blue,
  citecolor=mymagenta,
  menucolor=black,
  urlcolor=myblue,
  plainpages=false,
  pdfpagelabels,
  colorlinks=true,
  urlcolor=blue,
  anchorcolor = blue,
  hypertexnames=false
]{hyperref}

\newcommand{\heatcell}[1]{%
  \pgfmathsetmacro{\norm}{(#1 - 0.046)/(8 - 0.046)}%
  \pgfmathsetmacro{\norm}{min(1,max(0,\norm))}%
    \pgfmathtruncatemacro{\p}{100*\norm}%
    \edef\col{\noexpand\cellcolor{red!\p!white}}%
    \col #1%
}
\usepackage{cleveref}
\crefformat{figure}{Fig.~#2#1#3}
\crefformat{table}{Tab.~#2#1#3}
\crefformat{equation}{Eq.~(#2#1#3)}
\crefformat{section}{Sect.~#2#1#3}
\definecolor{mymagenta}{RGB}{200, 0, 100}
\definecolor{myblue}{RGB}{45, 48, 146}
\newcommand{\MeV}{\,\mathrm{MeV}}
\newcommand{\fm}{\,\mathrm{fm}}
\graphicspath{{plots/}}

\begin{document}
\title{
  The origin of excited states of the \texorpdfstring{$\Lambda$}{Lambda} baryon at the SU(3) point from Lattice QCD
  }
\author{Javier Suarez Sucunza}
\affiliation{Helmholtz-Institut für Strahlen- und Kernphysik (Theorie) and Bethe Center for Theoretical Physics, \\  Universität Bonn, 53115 Bonn, Germany}
\author{Thomas Luu}
\affiliation{Institute for Advanced Simulation 4 (IAS-4), Forschungszentrum Jülich, Germany}
\affiliation{Helmholtz-Institut für Strahlen- und Kernphysik (Theorie) and Bethe Center for Theoretical Physics,  Universit\"at Bonn, 53115 Bonn, Germany}
\author{Maxim Mai}
\affiliation{Albert Einstein Center for Fundamental Physics, Institute for Theoretical Physics, University of Bern, Sidlerstrasse 5, 3012 Bern, Switzerland}
\affiliation{The George Washington University, Washington, DC 20052, USA}
\author{Ferenc Pittler}
\affiliation{Computation-based Science and Technology Research Center,
  The Cyprus Institute, 20 Kavafi Str., Nicosia 2121, Cyprus}
\author{Carsten Urbach}
\affiliation{Helmholtz-Institut für Strahlen- und Kernphysik (Theorie) and Bethe Center for Theoretical Physics, \\  Universität Bonn, 53115 Bonn, Germany}
\affiliation{Bethe Center for Theoretical Physics, University of Bonn, Nussallee 12, 53115 Bonn, Germany}
\author{Haobo Yan (\begin{CJK*}{UTF8}{gbsn}燕浩波\end{CJK*})}
\affiliation{School of Physics, Peking University, Beijing 100871, China}

\begin{abstract}
  In this work we determine the finite-volume lattice QCD spectrum at the flavor symmetric $\mathrm{SU}(3)$ point in the meson-baryon singlet and octet irreducible representations. We construct the appropriate interpolation operators and perform the calculation on ensembles in quite large volume ($L=48$). We find three below-threshold energy levels, with the singlet having lower energy and the two octets being non-degenerate at one sigma, which for these large volumes ($M_{\pi} L\approx 14.5$) strongly suggests a bound state close to that energy at each of the irreducible representations.
  We confront this finite-volume spectrum with the prediction from UCHPT through the Lüscher method finding qualitative agreement. Finally we perform a re-fit of UCHPT free parameters to the available (experimental and lattice) data including the energy levels calculated in this work. This allows us to follow the pole trajectories to the physical point, identifying the $\Lambda(1405)$ as a lower octet, and $\Lambda(1380)$ as a singlet bound state in the $\mathrm{SU(3)}$ limit. Furthermore, $\Lambda(1670)$ is identified on a qualitative level as the heavier octet bound state and its relation to three-body final states is discussed.
\end{abstract}

\maketitle

\section{Introduction}

The $\Lambda(1405)$ and its internal structure have been a topic of
debate for a long time. Its mass is lighter than what would be
expected from the constituent quark model, while more modern
approaches based on Effective Field Theory also predict another state
with the same quantum numbers (now referred to as the $\Lambda(1380)$)
constituting a so-called two-pole structure~\cite{Oller:2000fj} (also
see Refs.~\cite{Mai:2020ltx,Meissner:2020khl,Hyodo:2020czb} for recent
reviews). In these Chiral Unitary (UCHPT) approaches the
$\Lambda(1405)$ is predicted to be a composite meson-baryon state, in
agreement with the most recent lattice and experimental
data~\cite{Pittler:2025upn}. The higher energy pole is recognized as a
4-star resonance in the PDG \cite{PDG} with spin parity $J^{P}=
\frac{1}{2}^{-}$ and strangeness $S= -1$ while the second
$\Lambda(1380)$ pole is still quoted only as a two-star resonance.

Experimentally, the $\Lambda(1405)$ was first discovered as a
resonance in the $\pi\Sigma$ channel in $\bar{K}N$ scattering just
below the $K^{-}p$ threshold~\cite{Dalitz:1959dq}. In fact, the
experimental investigation of the $\Lambda$ states is feasible for
$K^- p$ initial states, but not for instance for $\pi\Sigma$. 
Therefore, most of the experimental results are available above the
$\bar KN$ threshold such that analytic continuation to energies below
threshold is necessary to access the $\Lambda(1405)$ pole, and even
further for the $\Lambda(1380)$.
One way to circumvent this complication is to facilitate photon-induced
reactions~\cite{Niiyama:2009zza,CLAS:2013rjt,CLAS:2013rxx,CLAS:2014tbc,BGOOD:2021sog,J-PARCE31:2022plu}
and study more complex three-body final states (e.g.,
$K\pi\Sigma$). This allows one to scan the relevant energy regions
below threshold, but the theoretical treatment becomes more difficult,
see Refs.~\cite{Anisovich:2020lec,Roca:2015tea,Mai:2014xna,Lutz:2004sg}. 
Besides this, there are precise modern (sub-)measurements by the
SIDDHARTA~\cite{SIDDHARTA:2011dsy} and by the AMADEUS
collaborations~\cite{Piscicchia:2018rez,Piscicchia:2022wmd}. Notable
is also the recent progress on the kaonic deuterium experiment through the
SIDDHARTA-2. For a recent related review see
Ref.~\cite{Curceanu:2026zjg}.

While the theoretical investigation of baryons, and in particular
excited baryons, is similarly difficult, here the possibility of adjusting input quark masses offers
the opportunity to obtain information on these  $\Lambda$
states in possibly less demanding regions of the parameter space,
and eventually connect back to the physical point, where the quark
masses assume their physical values.
Specifically, at the $\mathrm{SU}(3)$-flavor symmetric point, the
additional symmetry changes the picture fundamentally, The
meson-baryon interaction can be decomposed into irreducible
representations (irreps) 
\begin{equation}
  \mathbf{8}_{B}\otimes\mathbf{8}_{M} =
  \mathbf{1}\oplus\mathbf{8}\oplus \mathbf{8'} \oplus\mathbf{10}
  \oplus\mathbf{\overline{10}} \oplus \mathbf{27}\, , 
\end{equation}
where the baryon octet is the one with positive parity. 
In the channels with the quantum numbers of the $\Lambda(1405)$ the
singlet and both octet representations are attractive which can lead
to bound states, e.g. see Refs.~\cite{Jido:2003cb,Bruns:2021krp,Guo:2023wes}. The decuplet
representations are non-interacting in these channels and the 27-plet
is repulsive. As one moves away from the $\mathrm{SU}(3)$ point the
poles corresponding to the bound states cannot disappear 
and move into the complex energy plane. It is, therefore, expected that
the two poles, one coming from the singlet, the other from one of the
octets, will move into the region of the $\Lambda(1405)$ giving way to
the two-pole structure at the physical point. Mapping out such a
trajectory from QCD is an important benchmark in the understanding of
the $\mathrm{SU}(3)$ hadron dynamics and the main motivation for the
current work.

In contrast to UCHPT, which predicts quark mass dependencies,
but not the values of the corresponding low-energy constants, Lattice
QCD is a first principles theoretical method, which ideally
complements UCHPT.
Lattice QCD provides an increasing number of constraints on the hadron
spectrum. For a recent review, see
Ref.~\cite{Mai:2022eur}. Investigations of meson-baryon scattering
have been performed previously~\cite{Torok:2009dg}, while the first
lattice studies of the $\Lambda(1405)$ looked at isolating the lowest
lying finite-volume spectrum using single baryon three-quark
operators~\cite{Gubler:2016viv,Menadue:2011pd, Engel:2012qp,
  Engel:2013ig, Nemoto:2003ft, Burch:2006cc,
  Takahashi:2009bu,Meinel:2021grq, Hall:2014uca}. More recently, there
have been studies using coupled-channel scattering analyses with
meson-baryon operators with ensembles near the physical
point~\cite{BaryonScatteringBaSc:2023zvt,BaryonScatteringBaSc:2023ori}.
A separate study, using the HAL QCD method, has investigated the $S$-wave
meson-baryon interaction in the singlet and octet channels in the
$\mathrm{SU}(3)$ limit~\cite{Murakami:2023phq,Murakami:2025oig}. In
this work we aim to provide a new input using --- for the first
time --- meson-baryon operators to calculate the energy levels of the
attractive irreps at the $\mathrm{SU}(3)$ point. We find that the
corresponding three energy levels lie below the non-interacting
threshold, the singlet being clearly distinct from the two octets, and
the two octet states being non-degenerate at the $1\sigma$ level.  

We use these new lattice results to perform an updated \emph{global} analysis
of the available experimental and lattice QCD data. 
For this, we utilize UCHPT at the next-to-leading chiral order relying
on the results of Ref.~\cite{Guo:2023wes,Pittler:2025upn}, see also
Refs.\cite{Lutz:2024ubv,Ren:2024frr}.
We map out the full chiral trajectory of all three states from the
$\mathrm{SU}(3)$-flavor symmetric to the physical point. We find that
the singlet state smoothly connects to the $\Lambda(1380)$, while the
lower/higher octet bound states connect to the $\Lambda(1405)$ and
$\Lambda(1670)$ resonance poles, respectively, at the physical point.


This paper is organized as follows: first, in \cref{section:group_theory}  we discuss the construction of the relevant $\mathrm{SU}(3)$ states and operator structures.  In \cref{section:lat_details} we then provide the details of the lattice simulation. In \cref{section:results} we show and discuss the obtained energy levels. Finally, in \cref{section:pole_trajectory} we introduce the Chiral Unitary Approach and the connection of the poles to the physical point.

\section{State and operator construction}
\label{section:group_theory}

The meson-baryon operators of interest contain four quarks and one anti-quark. The most general flavor wave function is
\begin{equation}
		\mathbf{\overline{3}}\otimes\mathbf{3}\otimes\mathbf{3}\otimes\mathbf{3}\otimes\mathbf{3}\,,
        \label{eq:full_vec_space}
\end{equation}
where they are all $\mathrm{SU}(3)$ flavor triplets. The vector space given by this tensor product is 243-dimensional, and can be decomposed into a direct sum of 21 irreducible representations, of which we are interested in the three that provide attractive interactions in the meson-baryon channels. We use the tensor formalism as explained in Ref.~\cite{Georgi:1999wka,Gregory:2021rgy,Gregory:2025ium} to calculate the states belonging to these irreps.  In this formalism the states that transform under the $\textbf{3}$ ($\mathbf{\overline{3}}$) irrep are represented as a vector with upper(lower) indices and the full vector space can be written as
\begin{equation}
  q_{i} p^{j} a^{k}b^{l}c^{m}\,.
\label{eq:vector_rep}
\end{equation}

To find the irreps, the general procedure consists in rewriting the tensor product in Eq.~\ref{eq:vector_rep} as a linear combination of terms, in which each one of them is totally symmetric in all upper indices, totally symmetric in all lower indices and traceless. We focus only on the relevant sectors for our calculation. For this we look at how the elements of the tensor product in Eq.~\ref{eq:full_vec_space} are combined to form the baryon and meson octets: 
\begin{equation}
		\begin{split}
          (\mathbf{\overline{3}}\otimes\mathbf{3})\otimes(\mathbf{3}\otimes\mathbf{3}\otimes\mathbf{3}) = (\mathbf{8}_{M}\oplus\mathbf{1})\otimes (\mathbf{10}\oplus\mathbf{8'}\oplus\mathbf{8}_{B}\oplus\mathbf{1})\,.
		\end{split}
\end{equation}
The meson octet appears in the combination of the anti-quark and one of the quarks, while the baryon octet appears in the combination of the remaining three quarks. We are interested in the tensor product of these two irreps and we know that the octet representations correspond to traceless tensors with one upper and one lower index, $u^{i}_{j}$. We can then obtain the projections into the irreps of interest by finding the irreps of the product of the baryon and meson octets, $(u_M)^{i}_{j}(v_B)^{m}_{h}$, and substituting into them the meson projectors,
\begin{equation}
	P[\mathbf{8}_{M}]^{j}_i = \left(q_ip^{j} -\frac{1}{3}\delta^{i}_j	q_h p^{h} \right) = u^{j}_i\,,
	\label{eq:mesonOctet}
\end{equation}
and baryon projectors,
\begin{equation}
\begin{aligned}
P[\mathbf{8}_{B}]^{klm} &= \frac{1}{2} \epsilon^{klh}\left(  
  \epsilon_{htf}a^{t}b^{f}c^{m}-\frac{\delta^{m}_{h}}{3}
  \epsilon_{etf}a^{t}b^{f}c^{e}\right) \\
&= \epsilon^{klh} v^{m}_{h}\,,
\end{aligned}
\label{eq:baryonOctet}
\end{equation}
respectively.

The decomposition of the tensor product of baryon and meson octets yields
\begin{equation}
  P[\mathbf{1}]_i^{jklm} =A\delta^{j}_h \delta^{m}_i \epsilon^{klh}u^{\alpha}_\beta v^{\beta}_\alpha 
\end{equation}
for the singlet,
\begin{equation}
P[\mathbf{8}]_i^{jklm} = B\delta^{j}_h \epsilon^{klh} \left(u^{\alpha}_i v^{m}_\alpha -\frac{1}{3}\delta^{m}_{i}u^{\alpha}_{\beta}v^{\beta}_{\alpha}\right)
\end{equation}
for the first octet, and
\begin{equation}
 P[\mathbf{8'}]_i^{jklm} = C\delta^{m}_i \epsilon^{klh} \left(u^{j}_\alpha v^{\alpha}_h -\frac{1}{3} \delta^{j}_h u^{\beta}_\alpha v^{\alpha}_\beta\right)  
\end{equation}
for the second octet. $A$, $B$ and $C$ are overall constants in each projector that are not relevant because after projecting we can normalize the vectors.

The singlet and the octet irreps have different quantum numbers, so we expect no mixing between them. However, the two octet representations have the same quantum numbers, so there is nothing preventing their mixing. Physically this means that both octet irreducible representations will couple to the same physical states. In UCHPT  the interaction is diagonalized to isolate the physical states \cite{Bruns:2021krp}. In our case, the construction of the two irreps is such that the Wick contraction of states from the two different octet irreps is zero. Within each octet we have that
\begin{equation}
  \contraction[1ex]{\bigl\langle\,}{\mathbf{8}_{i}}{}{\overline{\mathbf{8}_{j}}}
    \left\langle \mathbf{8}_i  \overline{\mathbf{8}_j} \right\rangle \sim \delta_{ij} \,,
\end{equation}
where $i$ and $j$ indicate the state within each irreducible representation. Within each irreducible representation, the contraction of any state with itself gives the same result as expected at the $\mathrm{SU}(3)$ point.

For each irreducible representation we have used a basis of three bilocal interpolation operators
\begin{equation}
\label{eq:op_basis}
\begin{aligned}
\mathcal{O}_1 & \sim \gamma_5 P_{+} q_1 (q_2 C \gamma_5 q_3) (q_4 \gamma_5 \bar{q_5})\,,\\
\mathcal{O}_2 & \sim \gamma_5 P_{+} \gamma_5 q_1 (q_2 C q_3) (q_4 \gamma_5 \bar{q_5})\,, \\
\mathcal{O}_3 & \sim  \gamma_5 P_{+} q_1 (q_2 i \gamma_4 C \gamma_5 q_3) (q_4 \gamma_5 \bar{q_5})\,, \\
\end{aligned}
\end{equation}
where the meson is interpolated by the pseudoscalar operator with negative parity and zero spin; and the baryon is interpolated by three operators, all with spin one-half and projected into positive parity. The extra $\gamma_5$ is added to ensure the whole operator has negative parity. The operators are calculated at zero total momentum and zero relative momentum between baryon and meson operators, therefore they belong to the $G_{1,u}(0)$ irrep. We did not include single-hadron operators because they would introduce mixing between the $\mathbf{8}$ and $\mathbf{8'}$ operators; their inclusion is left for a future publication. 

\section{Lattice Action and Analysis}
\label{section:lat_details}

Our measurements are performed on one ensemble (denoted as C103) of gauge configurations generated by the CLS collaboration~\cite{Horz:2020zvv,BaryonScattering:2025ziz}. This ensemble was generated with a Clover-Wilson action with $N_f = 2+1$ dynamical fermions with $m_u=m_d=m_s$. The lattice spacing is $a\approx 0.086\fm$, $V=48^{3}\times 96$ and $M_\pi L \approx 14.5$ which is very large and allows us to have the exponentially suppressed finite volume effects under control. In total for this work we have used 800 gauge configurations coming from two different replicas with 400 each.

In the calculation of the correlation functions, the presence of quark and antiquark operators of the same flavor in the same time slice requires the evaluation of all-to-all propagators. For their calculation we use the well-established method of distillation~\cite{HadronSpectrum:2009krc}. Distillation is a smearing of the quark fields using a truncated space of eigenvectors of the lattice Laplace operator. The truncation is possible because the higher modes are exponentially suppressed. For the calculations in this work we used a Laplacian subspace spanned by $N_e = 100$ eigenvectors. 

For the UCHPT calculations the pion decay constant, $f_\mathrm{PS}$, is needed. Since we are in the $\mathrm{SU}(3)$ limit the decay constant will be the same for all pseudoscalar mesons. We have calculated it following Ref.~\cite{Bruno:2016plf} without using the improved axial current operator
\begin{equation}
  f_{\mathrm{PS}} = Z_{A}(g_0) f_{\mathrm{PS}}^{\text{bare}}\,,
\end{equation}
with the renormalization factor, $Z_{A}(g_0) =0.75629(65)$, taken from Ref.~\cite{DallaBrida:2018tpn}. The bare decay constant is calculated from axial and pseudoscalar matrix elements as described in section 3 in Ref.~\cite{Kuberski:2024pms} with a value of $f_{\mathrm{PS}^{\mathrm{bare}}}= 267(13)\MeV$. The value of the renormalized decay constant is $f_{\mathrm{PS}} = 202.3(9.7)$ MeV, using the convention in which $f_{\pi}$ at the physical point is $\sim130 \MeV$. In \cref{table:latt_params} we include a summary table of the relevant physical quantities on the C103 ensemble. There and throughout the article the meson/baryon masses are denoted by $M/m$, respectively. 

The statistical uncertainties were estimated using the bootstrap resampling procedure. The calculation of the Laplacian eigenvectors and perambulators were performed with PyQuda~\cite{Jiang:2024lto}. The currents necessary for the calculation of the decay constant were computed with chroma~\cite{Edwards:2004sx}. In both cases the inversions were performed using the QUDA library~\cite{Clark:2009wm}.

\begin{table}[t]
\centering
\caption{Lattice quantities measured in the C103 ensemble.}
\label{table:latt_params}

\setlength{\tabcolsep}{6pt}
\renewcommand{\arraystretch}{1}

\begin{tabular}{ccccc}
\hline
$a/\mathrm{fm}$ & $L^3\times T$ & $M/\MeV$ & $m/\MeV$ & $f_{\mathrm{PS}}/\MeV$ \\
\hline
0.086 & $48^3\times 96$ & 713.42(32) & 1613.8(2.2) & 202.3(9.7) \\
\hline
\end{tabular}
\end{table}

\section{Results}
\label{section:results}

We extract the lowest lying energy levels from the correlator matrix defined in \cref{eq:op_basis} by solving the GEVP \cite{Michael:1982gb,Luscher:1990ck,Blossier:2009kd} for each irrep
\begin{equation}
  C^{(t)}v^{(n)} = \lambda^{(n)} C(t_0) v^{(n)}(t)\,,
\end{equation}
the eigenvalues (principal correlators) decay exponentially in $t$ like
\begin{equation}
  \lambda_n(t) \sim e^{-E_n(t)}(1+\mathcal{O}(e^{-\Delta E_n (t-t_0)}))\,,
\end{equation}
where $E_n$ is the nth-state energy and $\Delta E_{n}$ is the energy difference to the next excited state. The energy levels were extracted by fitting a constant to the effective mass data
\begin{equation}
  m_\mathrm{eff}(t) = \log \left(\frac{\lambda_n(t)}{\lambda_n(t+1)} \right)\,,
\end{equation}
which approaches $E_n$ for large enough Euclidean times $t$. We used three by three matrices containing all operators for the $\mathbf{1}$ and $\mathbf{8}$ irreps. For the $\mathbf{8}$' we only used  $\mathcal{O}_1$ and $\mathcal{O}_2$, because we found that the third operator only increased the noise without improving the signal. 

\begin{figure}[t]
  \centering
  \includegraphics[width=\linewidth]{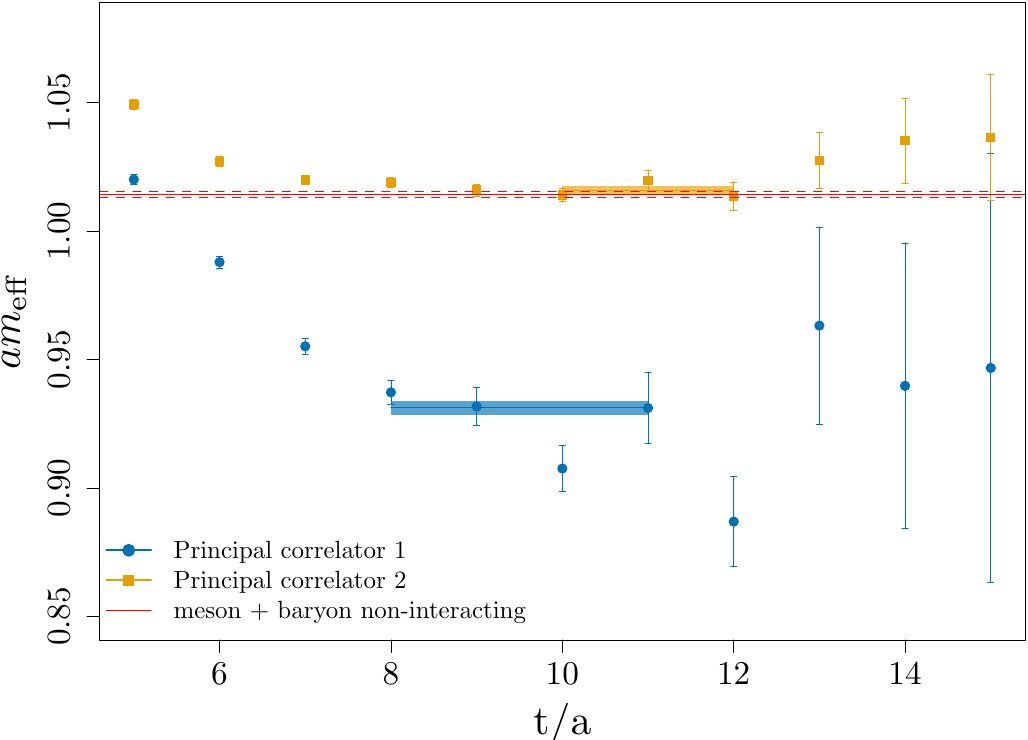}
  \caption{Effective mass of the principal correlators from the GEVP for the singlet in lattice units. The red line indicates the non-interacting meson-baryon threshold.}
  \label{fig:gevp_singlet}
\end{figure}

In \cref{fig:gevp_singlet} we show the effective mass of the principal correlators for the singlet. The lowest lying state is below the two-particle threshold indicating an attractive interaction, which together with the high $M_{\pi}L$ of our ensembles, suggests that it may be a bound state. The first excited state sits at the meson-baryon threshold energy. It is interesting that its effective mass has smaller errors and a better signal than the ground state. The reason is that, because of the intrinsic meson-baryon structure of the interpolation operators, there is a very large overlap with the non-interacting meson-baryon states on the lattice, even more than with the interacting state. For this reason, the inclusion of a basis of operators was crucial to resolve the ground state, in particular that of the operator $\mathcal{O}_2$. 

\begin{figure}[t]
  \centering
  \includegraphics[width=0.7\linewidth]{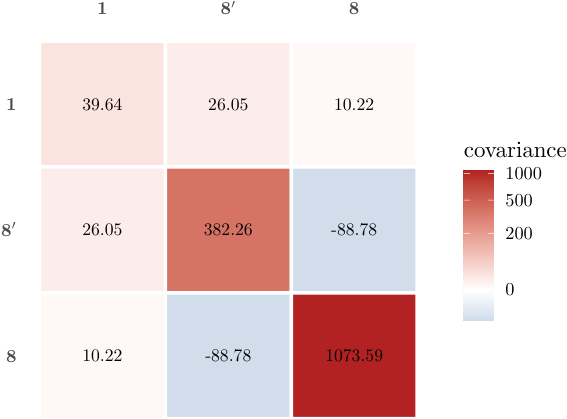}
  \caption{Covariance matrix of the ground state energy levels computed from the bootstrap samples.}
  \label{fig:covariance_matrix}
\end{figure}

\begin{figure*}[!htbp]
\centering
\begin{minipage}{0.48\textwidth}
  \centering
  \includegraphics[width=\linewidth, page=1]{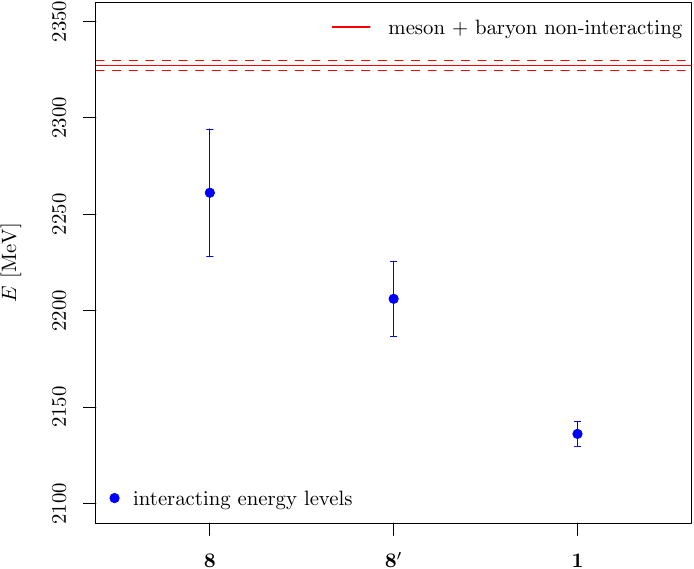}
\end{minipage}\hfill
\begin{minipage}{0.48\textwidth}
  \centering
  \includegraphics[width=\linewidth, page=2]{plots/e_levels}
\end{minipage}
\caption{\textbf{Left}: Energy levels $E$ in MeV for the singlet
  $\mathbf{1}$,  octet $\mathbf{8}$ and octet
  prime $\mathbf{8}'$ irreps. The red line indicates the
  non-interacting meson-baryon
  threshold. \textbf{Right}: Correlated difference $\Delta E$ between
  the energy levels for the three combinations of the three irreps
  $\mathbf{1}, \mathbf{8}$ and $\mathbf{8}'$.}
  \label{fig:energy_levels}
\end{figure*}

We have repeated the same calculation for the two octet states. Since
the qualitative behaviour is the same, we leave the effective mass
plots for the appendix \ref{section:ap_gevp} and directly show the
energy levels in the left panel of \cref{fig:energy_levels}, and quote them here
\begin{align}
  E_{1} &= 2136.2(6.3) \MeV\,,\nonumber\\
  E_{8'}&= 2206(19) \MeV\,, \label{eq:SU3:fin-vol-spectrum} \\ 
  E_{8} &= 2261(33) \MeV\,.\nonumber
\end{align}
We also provide the covariance matrix in \cref{fig:covariance_matrix}.

It appears that the energies for all three irreps are below the meson-baryon threshold (2327.2(2.5) MeV), and we consider them bound states as well. While the
$\mathbf{1}$ and $\mathbf{8}'$ states are statistically significant below that
threshold, the octet is only $2\sigma$ below threshold.
In general, we observe that the singlet is lighter in mass than both
octet states ($E_\mathbf{1} < E_\mathbf{8} , E_{\mathbf{8}'}$ ), in agreement with previous
lattice results~\cite{Murakami:2025oig} and with the predictions of
UCHPT \cite{Jido:2003cb}.

To better understand the energy splitting between the separate states, we show the energy differences between all three distinct pairs of irreps in the right panel of \cref{fig:energy_levels}, facilitating the statistical correlation between the energy levels. They read
\begin{equation}
\begin{split}
  \Delta E_{\mathbf{1}-\mathbf{8}'} &= -69(19) \MeV\,,\\
  \Delta E_{\mathbf{1}-\mathbf{8}}  &= -125(33) \MeV\,, \\
  \Delta E_{\mathbf{8}'-\mathbf{8}} &= -56(40) \MeV\,.
\end{split}
\end{equation}
We observe a clear indication with more than $2\sigma$ confidence for the singlet state to be non-degenerate with both octet states. Also the octet-octet' energy splitting is non-zero within errors, however, only barely.

We have also calculated the $q\cot(\delta)$ of these energy levels
\begin{equation}
  q \cot(\delta( q))= \frac{2\,\mathcal{Z}_{00}(1;q^2)}{\sqrt{\pi}L}
\end{equation}
where $E = \sqrt{m_1^2 + q^2} + \sqrt{m_2^2 + q^2}$ and \(\mathcal{Z}_{00}\) is the Lüscher zeta function~\cite{Luscher:1986pf,Luscher:1990ux,Luu:2011ep}. We use the convention in which an attractive $a_0$ is negative. We show the results in \cref{tab:q2_qcotd}.

The uncertainties for $q\cot(\delta(q))$ were calculated by propagating the bootstrap samples and applying an outlier removal, removing data points that are outside the range $[Q_{16}-1.5\text{IPR},Q_{86}+1.5\text{IPR}]\,$, where $\text{IPR}= Q_{84}-Q_{16}$ is the central 68\% interpercentile range. This is necessary because the energy levels are close to threshold and the bootstrap distribution is distorted by the singularity of the Lüscher zeta function. 

\begin{table}[b]
\centering
\caption{Results for $q^2$ and $q\cot(\delta)$ in the different $\mathrm{SU(3)}$ channels.}
\label{tab:q2_qcotd}

\setlength{\tabcolsep}{19pt}
\renewcommand{\arraystretch}{1}

\begin{tabular}{|c|c|c|}
\hline
\rule{0pt}{3ex}Channel & $q^2/\MeV^2$ & $q\cot(\delta)\times \fm$ \\
\hline
\textbf{1}    & $-1.939(59)$ & $-2.119(32)$ \\
\textbf{8}$'$ & $-1.27(19)$  & $-1.71(13)$  \\
\textbf{8}    & $-0.69(34)$  & $-1.26(35)$  \\
\hline
\end{tabular}
\end{table}

\section{Connecting trajectory to the observable world}
\label{section:pole_trajectory}

\subsection{Recap of the Chiral Unitary Approach}

The chiral symmetry of QCD provides a model independent way to construct an effective Lagrangian formalism in terms of hadronic degrees of freedom, allowing for a systematically improvable approach to hadronic interactions at low-energies. Extended to the meson-baryon sector this is the essence of Baryon Chiral Perturbation Theory (CHPT) or (BCHPT)~\cite{Gasser:1987rb, Bernard:1992qa, Tang:1996ca, Ellis:1997kc}. In its minimal form, such a Lagrangian has been so far scrutinized up to next-to-next-to-leading ($p^3$) order~\cite{Frink:2006hx} (see also Refs.~\cite{Holmberg:2018dtv,Song:2024fae}).  
When approaching meson-baryon scattering from perturbative CHPT calculations, one faces several roadblocks. First, the low-energy constants, with their abundance exponentially growing with the chiral order, are quite poorly known. For a recent lattice based determination see, e.g., Ref.~\cite{Lutz:2023xpi} and references therein. Second, the convergence of that chiral series can be hindered by large meson masses and involved momenta as it is the case for antikaon-nucleon scattering. For an explicit calculation see Ref.~\cite{Mai:2009ce}. Furthermore, the presence of the $\Lambda(1405)$ in the isoscalar part of such an amplitude just below the $\bar KN$ threshold renders any perturbative expansion meaningless. These challenges have motivated a substantial number of works combining constraints from CHPT with unitarization techniques, leading to the so-called chiral unitary approaches (UCHPT). For example applications see Refs.~\cite{Oller:2000fj,Lutz:2001yb, Borasoy:2005ie, Oller:2005ig,Oller:2006jw,Hyodo:2008xr, Ikeda:2012au, Mai:2012dt, Guo:2012vv, Mai:2014xna,Ramos:2016odk,Kamiya:2016jqc,Cieply:2016jby, Sadasivan:2018jig,Lu:2018zof,Oller:2019opk, Feijoo:2021zau, Sadasivan:2022srs,Lutz:2024ubv} and for a comparison of various models see Ref.~\cite{Mai:2020ltx}.

In this work, we rely on the results of the recent UCHPT study~\cite{Pittler:2025upn}, applied to describe simultaneously all available experimental scattering and threshold data (physical point) as well as the finite-volume energy eigenvalues from a recent lattice QCD calculation of the ($I=0, S=-1$) meson-baryon system by the BaSC collaboration at $M_\pi\approx 200\MeV$ (BaSC point). For the latter the multichannel Lüscher method was applied with respect to the UCHPT $K$-matrices. 

In that study, the dynamics of the meson-baryon system is determined through the on-shell reduced form of the $T$-matrix
\begin{align}
    T(s)&=-V(s)+T(s)G(s)V(s)=\frac{-V(s)}{1-V(s)G(s)}\,.
\label{eq:T-matrix}
\end{align}
This fulfills two-body unitarity, being expressed through a chiral potential $V$ and a meson-baryon loop integral $G$, which are matrices with respect to all possible two-body channels built from the ground state octet of pseudoscalar mesons and baryons. Specifically, the available channels are ${\cal S}=\{K^-p$, $\bar K^0 n$, $\pi^0\Lambda$, $\pi^0\Sigma^0$, $\pi^+\Sigma^-$, $\pi^-\Sigma^+$,$\eta\Lambda$, $\eta \Sigma^0$, $K^+\Xi^-$, $K^0\Xi^0\}$. 

The chiral order to which the above expression matches the perturbative chiral series is determined through that of the chiral potential $V$. In Ref.~\cite{Pittler:2025upn}, three choices are considered
\begin{align}
    &V_{\alpha\beta}=
    \underbrace{\underbrace{\underbrace{V_{\alpha\beta}^{\mathrm{WT}}}_{\text{M1}} +V_{\alpha\beta}^{\mathrm{BORN}s}+V_{\alpha\beta}^{\mathrm{BORN}u}}_{\text{M2}}+V_{\alpha\beta}^{\rm{NLO}}}_{\text{M3}}\,.
\label{eq:VNLO}
\end{align}
The unitarization procedure comes with a price of introducing a certain model dependence. Among others this is reflected in residual dependence on the regularization procedure, resulting for instance in a choice of subtractions ($a_{\alpha\in{\cal S}}$) entering the dimensionally regularized log-divergent meson-baryon loop integral
\begin{align}
\label{eq:G_loop}
    &G_\alpha(\sqrt{s}) =
    a_\alpha+
    \\
    \nonumber
    &\frac{1}{32\pi^2}\left(
      \log\left(\frac{m_\alpha^2}{\mu^2}\right)
      +\log\left(\frac{M_\alpha^2}{\mu^2}\right)\right.-\frac{m_\alpha^2-M_\alpha^2}{s}\log\left(\frac{M_\alpha^2}{m_\alpha^2}\right)\\\nonumber
    &\left.
      ~~~~~~~~~~~~~~~~
      -2-\frac{8p_\alpha}{\sqrt{s}}\operatorname{arctanh}\left(\frac{2\sqrt{s} p_\alpha}{(m_\alpha+M_\alpha)^2-s}\right)
      \right)\,.
\end{align}
Three examples are considered in Ref.~\cite{Pittler:2025upn} to map out this source of model dependence, which are taken as a starting point of this study as well.

\subsection{UCHPT prediction and comparison to lattice result}

To make a comparison with the obtained finite-volume spectrum at the $\mathrm{SU}(3)$ point we use the well-established Lüscher method extended to coupled-channels and related to the $K$-matrix from the UCHPT amplitude~\cref{eq:T-matrix} as 
\begin{align}
  \label{eq:LueschersMethod}
  &\operatorname{det}(1-\tilde K(s)B(s,L))=0\,.
\end{align}
Here, the $B$-matrix is a geometric function of box-size $L$ only, see Ref.~\cite{Pittler:2025upn} for explicit relations. 

To access the particular irrep of the increased $\mathrm{SU}(3)$ symmetry we start with the isospin projected $4\times4$ $K$-matrix and project it to {singlet, octet, 27plet} through the procedure outlined in Ref.~\cite{Bruns:2021krp} (see also Refs.~\cite{Kamiya:2016jqc, Guo:2023wes, Lu:2024ajt}) reading
\begin{align}
  &K_{I=0}\longrightarrow K_{SU3}:=P^\dagger K_{I=0} P\,,\\
  \quad
  &P=\left(
    \begin{array}{cccc}
    -\sqrt{\frac{3}{8}} & \sqrt{\frac{3}{5}} & 0 & -\frac{1}{\sqrt{40}} \\
    -\frac{1}{2} & -\frac{1}{\sqrt{10}} & \frac{1}{\sqrt{2}} & \sqrt{\frac{3}{20}} \\
    \frac{1}{2 \sqrt{2}} & \frac{1}{\sqrt{5}} & 0 & \sqrt{\frac{27}{40}} \\
    \frac{1}{2} & \frac{1}{\sqrt{10}} & \frac{1}{\sqrt{2}} & -\sqrt{\frac{3}{20}} \\
    \end{array}
  \right)\,.
\end{align}
As discussed there, both octets mix in a certain way such that the $2\times 2$ octet matrix needs to be diagonalized for each set of parameters and kinematic variables. To our knowledge this cannot be overcome by projections discussed in previous chapters, presumably due to different field normalizations in CHPT and on the lattice. For this reason, we chose to call both octet states determined in the UCHPT procedure as $\mathbf{8}$a and $\mathbf{8}$b (ordered from low to high energy), differentiating from the previous $\mathbf{8}$, $\mathbf{8}'$ notation.

\begin{figure}[t]
  \centering
  \includegraphics[width=0.48\textwidth]{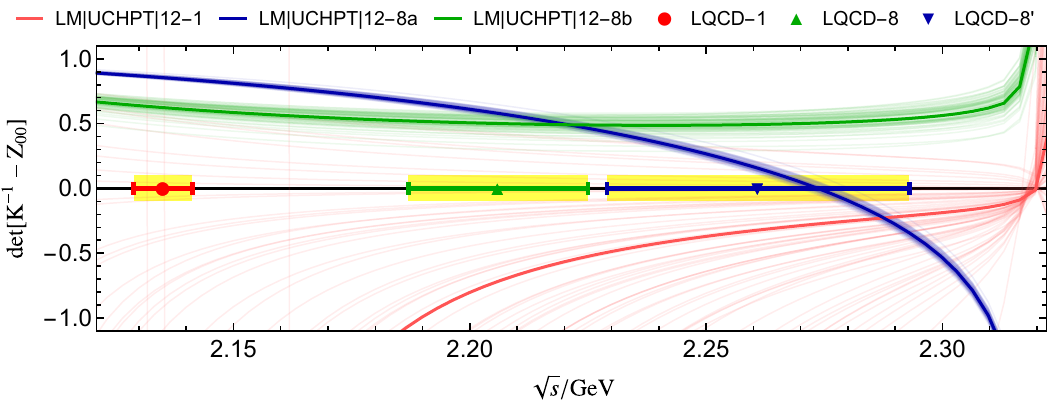}
  \includegraphics[width=0.48\textwidth]{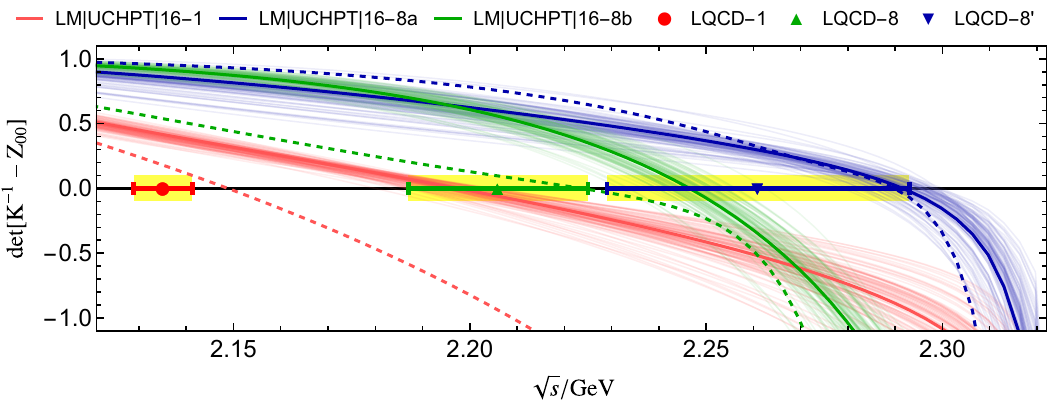}
  \caption{Predicted UCHPT energy eigenvalues in the octet and singlet irreps, determined as zeros of $\operatorname{det}[K^{-1}-Z_{00}]$ according to the Lüscher method. Only the most data-adequate next-to-leading models (fit $F_{12}$ and $F_{16}$) are shown and compared with the Lattice results. The lower panel includes the result (dashed line) of a refit of the solution $F_{16}$, i.e. $F_{16'}$.
  \label{fig:SU3pred16}}
\end{figure}
Predictions of our best models (fit $F_{12}$ and $F_{16}$ from Ref.~\cite{Pittler:2025upn}) are presented and compared with the Lattice results in \cref{fig:SU3pred16}. They constitute the most sophisticated models (type M3 in \cref{eq:VNLO}) leading to best Akaike information criterion (AIC) and Bayesian information criterion (BIC) values together with the best data description. Statistical uncertainties determined through a re-sampling procedure lead to a spread of predicted values as shown in that figure as well. We note that fit 12 agrees only with one octet eigenvalue, missing the other two significantly. One explanation for this disagreement is the specific form of the subtraction procedure of this type of model, which includes an additional dimensionful quantity $\Lambda$ (matching cutoff). The unknown and thus neglected quark mass dependence of that quantity allowed connecting (BaSC point) with the physical one, but may not be valid when extrapolated to a more distant $\mathrm{SU}(3)$ point. On the contrary, there is no such quantity in the scheme leading to the fit $F_{16}$, which predicts a spectrum very similar to the one determined on the lattice (see bottom panel of \cref{fig:SU3pred16}). Both octet eigenvalues agree within 2 standard deviations. The pattern of all three states also agrees, while a disagreement of a few standard deviations for the singlet state emphasizes the precision of the determined lattice eigenvalues. Crucially, this shows that this Lattice input can, indeed, provide new constraints on the scattering amplitudes.

\begin{figure}[b]
  \centering
  \includegraphics[width=0.48\textwidth]{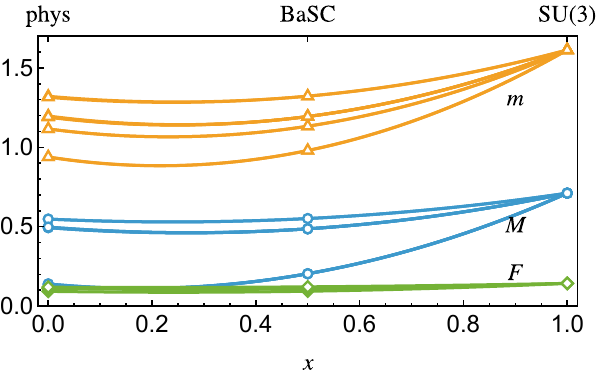}
  \caption{Assumed trajectory of dimensionful quantities $\aleph(x)=(F,M,m)$ required for evaluation of the UCHPT amplitude. Heuristic parameter $x$ scales an exact connection between the physical, BaSC and $\mathrm{SU}(3)$ point.
  \label{fig:trajectory}}
\end{figure}

Staying with the best fit $F_{16}$ which appears to be in qualitative agreement with the $\mathrm{SU}(3)$ results we wish to connect them to see if and how the measured states relate to the analytic structure of the UCHPT scattering amplitude and resonance poles. For this, we define a smooth trajectory for meson decay constants, meson and baryon masses $\aleph=(F,M,m)$ through a heuristic quadratic function of a parameter $x\in[0,1]$. In that, we demand for $\aleph(x)$ that $\aleph(0.0)=\aleph_{phys}$, $\aleph(0.5)=\aleph_{BaSC}$, $\aleph(1.0)=\aleph_{\mathrm{SU}(3)}$, which fixes $\aleph(x)$ completely as depicted in \cref{fig:trajectory}. We emphasize again that this is not a QCD trajectory but rather a choice to smoothly and exactly connect three distinct points.

\begin{figure*}[t]
  \centering
  \includegraphics[width=0.99\textwidth]{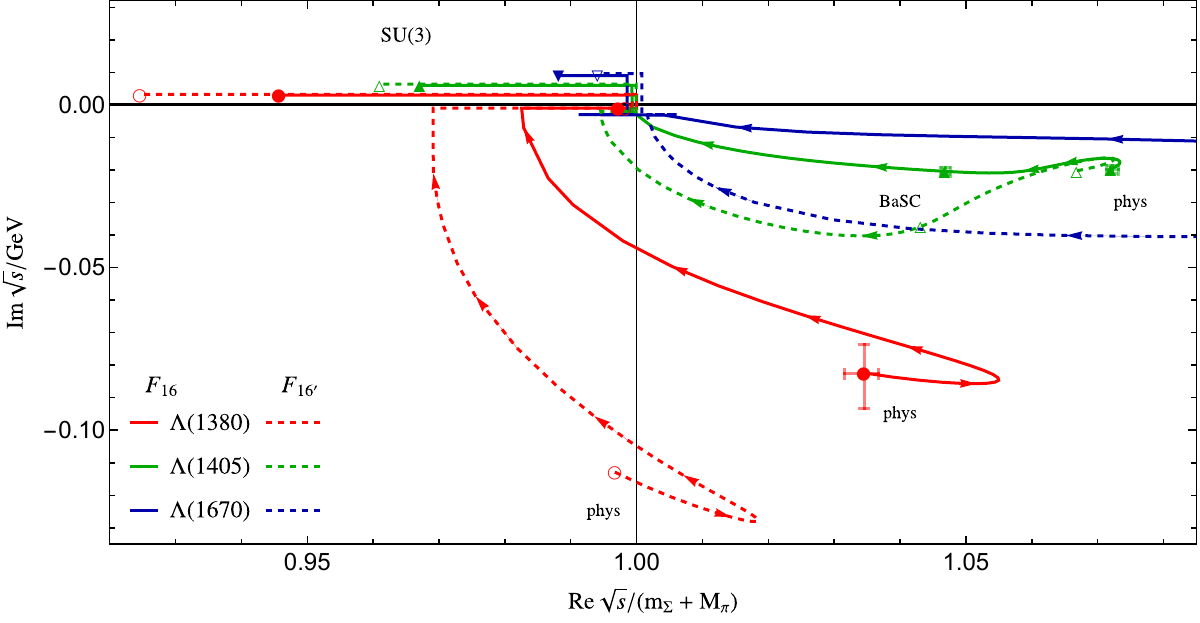}
  \caption{Mass (chiral) trajectory of the pole positions of isoscalar excited $\Lambda$ states. Full and dashed lines denote the pole positions as a function of $x\in[0,1]$ for the fit $F_{16}$ and $F_{16'}$, respectively. The former includes statistical uncertainties being fitted to the data at the physical, BaSC ($M_\pi\approx 200\MeV$), while the former also contains the input from the finite-volume lattice QCD spectrum at the $\mathrm{SU}(3)$ point. Symbols depict the pole positions at that input points while the arrows are put to guide the eye. Small arbitrary positive (negative) offset is added to the pole positions when these become (virtual) bound states on the physical and physical sheets, respectively.
  \label{fig:CelestialDance} 
  }
\end{figure*}
Observing the pole trajectory on the complex energy plane at the $\mathrm{SU}(3)$ point --- depicted in \cref{fig:CelestialDance} --- we identify three bound state poles, each of which leading to one energy eigenvalue. The explicit dd depiction can be found in \cref{appendix:poles}. Following the trajectory to the physical point, the singlet bound state becomes less bound, becoming then a virtual bound state at the BaSC point ($x=0.5$). It then goes into complex plane (Riemann sheet $[++----++++]$ with respect to the two-body branching points $\cal{S}$) where it ends as the $\Lambda(1380)$ pole at the physical point ($x=0$). The deeper bound octet state undergoes a similar transformation becoming, however, a virtual state at $x\approx 0.8$, connecting smoothly then to the $\Lambda(1405)$ pole (Riemann sheet $[++----++++]$) at the physical point. Finally, the shallow octet state becomes a virtual state at $x\approx 0.95$ and finally ends as a resonance pole of the $\Lambda(1670)$ at the physical point on the sheet $[-------+++]$. We note here that the experimental input included in the fits of Ref.~\cite{Pittler:2025upn} did not extend in that energy range. Therefore, we have no expectation that the extracted pole position needs to agree with the data-driven determinations quoted by the PDG~\cite{PDG}.

\subsection{UCHPT re-fit at SU(3), BaSC and physical point}

As a final step we have performed a re-fit of all available experimental and lattice inputs, including now also the finite-volume spectrum at the $\mathrm{SU}(3)$ symmetric point~\cref{eq:SU3:fin-vol-spectrum}. In that, one is forced to compare old bubble chamber data from the 1960s with few but very precise modern experimental or numerical (lattice QCD) results. Weighting all inputs equally may introduce a bias towards old and abundant data with largely unknown systematics. Thus, and as further discussed in Ref.~\cite{Mai:2020ltx} we minimize 
\begin{align}
    \chi^2_{\rm dof}=\frac{\sum_a N_a}{A((\sum_a N_a)-N_{\rm par})}
    \sum_{a=1}^{A}\frac{\chi^2_a}{N_a}\,.
    \label{eq:chi2}
\end{align}
In the present case number of ``measured'' quantities is $A=11$, number of free parameters (low-energy constants) $N_{\rm par}=7$,  number of data for each data type $a$ is $N_a$. The total number of data is denoted by $N_{\rm data}=\sum_a N_a$. Note that all correlations between the data are included when available. For more details of the fit procedure see Ref.~\cite{Pittler:2025upn}. 

Starting from the fit $F_{16}$ of Ref.~\cite{Pittler:2025upn} which qualitatively describes the new lattice input, we perform a re-fit $F_{16'}$. Indeed, we obtain a good description of all inputs with a total $\chi^2_{\rm dof,16'}\approx 2.71$. This is quite similar to the fit quality the fit $F_{16}$ without the $\mathrm{SU}(3)$ data, i.e., $\chi^2_{\rm dof, 16}\approx 2.12$. Taking a look on the separated contributions in \cref{tab:chi2-refit} we note that the improvement of the description at the $\mathrm{SU}(3)$ point comes with a price of worse description of the BaSC finite-volume spectrum. Whether there is a real conflict between the data, one cannot say at this point. One particularly interesting point in this regard is the scale setting relevant for the extraction of hadron parameters, see, e.g., \cite{Hu:2017wli,Mai:2019pqr,Lutz:2024ubv,Hudspith:2024kzk}. This issue together with a more in-depth discussion of the statistics weights \cref{eq:chi2} are the main reasons for not following up on statistical uncertainty for this re-fit. Overall, we note that the new $\mathrm{SU}(3)$ input, while few, is quite constraining. 
\newcommand{\vmin}{0.046}
\newcommand{\vmax}{4.866}
\begin{table}[b]
  \caption{Unified description of lattice $\mathrm{SU}(3)$ [this work], lattice BaSC~\cite{BaryonScattering:2025ziz,BaryonScatteringBaSc:2023ori,BaryonScatteringBaSc:2023zvt} and the experimental input through the UCHPT model~\cite{Pittler:2025upn}. Asterisk marks the contribution of not fitted quantities, colors are put for convenience to mark the biggest contribution (red).
  \label{tab:chi2-refit}}
  \begin{tabular}{|l|c|c|c|}
    \hline
    &$N_a$& $\chi^2_a(F_{16})$ & $\chi^2_a(F_{16}^{\prime})$ \\
    \hline
    AMADEUS                                 &1  & \heatcell{0.13} & \heatcell{0.36} \\
    SIDDHARTA \& Threshold ratios           &5  & \heatcell{2.23} & \heatcell{0.05} \\
    $\sigma_{K^{-}p\to K^{-}p}$             &83 & \heatcell{2.05} & \heatcell{4.87} \\
    $\sigma_{K^{-}p\to {\bar{K}}^{0}n}$     &47 & \heatcell{3.92} & \heatcell{4.34} \\
    $\sigma_{K^{-}p\to \pi^{0}\Lambda}$     &11 & \heatcell{3.00} & \heatcell{3.87} \\
    $\sigma_{K^{-}p\to \pi^{0}\Sigma^{0}}$  &11 & \heatcell{1.66} & \heatcell{0.80} \\
    $\sigma_{K^{-}p\to \pi^{+}\Sigma^{-}}$  &51 & \heatcell{3.35} & \heatcell{1.96} \\
    $\sigma_{K^{-}p\to \pi^{-}\Sigma^{+}}$  &49 & \heatcell{1.90} & \heatcell{3.13} \\
    Energy eigenvalues LQCD${}_{\rm BaSC}$  &14 & \heatcell{1.73} & \heatcell{3.40} \\
    $m_{\mathrm{BaSC}}$                     &4  & \heatcell{0.68} & \heatcell{2.11} \\
    Energy eigenvalues LQCD${}_{\mathrm{SU}(3)}$ &3   & \heatcell{33.28}${}^*$ & \heatcell{1.82} \\
    \hline
  \end{tabular}
\end{table}

Taking a look on the chiral trajectory of the resonance poles from the updated fit $F_{16'}$, dashed lines in \cref{fig:CelestialDance} we observe large qualitative similarities with the previous solution (full line). For example, all three poles of $\Lambda(1380)$, $\Lambda(1405)$ and $\Lambda(1670)$ follow similar transmutation for 
\begin{center}
  resonance$ \leftrightarrow$ virtual~bound~state $\leftrightarrow$ bound state\\
  $0\underset{x}{\longleftrightarrow}1$
\end{center} 
as described above. Specifically, the depiction of the pole structure at the $\mathrm{SU}(3)$ point can be found in \cref{appendix:poles}. Quantitatively, the pole position of the $\Lambda(1380)$ differs beyond expected statistical uncertainty at the physical and $\mathrm{SU}(3)$ points. The pole positions of the $\Lambda(1405)$ are closer to each other at the physical point but deviate from each other at the $\mathrm{SU}(3)$ and the BaSC points. Trajectories of $\Lambda(1670)$ do also vary between $F_{16}$ and $F_{16'}$ solutions, however, at the $\mathrm{SU}(3)$ point both are quite close to each other. Thus, one can consider this as mostly predicted from chiral symmetry and $\mathrm{SU}(3)$ lattice QCD results as no experimental input is taken into account at energies in the region of that resonance. An interesting aspect in this regard is that at the physical and also the BaSC point this state is well above the lowest three-body threshold ($\pi\pi\Lambda$), see. e.g. Fig. 5 in Ref.~\cite{Pittler:2025upn}. Thus, a correct description of that state needs to take the three-body unitarity into account --- possibly through the recently developed formalism~\cite{Mai:2017vot,Mai:2017bge}, applied to a variety of coupled-channel systems~\cite{Mai:2017vot,Mai:2017bge,Doring:2025phq,Feng:2024wyg,Yan:2025mdm,Feng:2026ixm}. This highlights another advantage of the lattice simulations at the $\mathrm{SU}(3)$ point, or rather high pion mass. At this point the relative position of the three-particle threshold moves above the pole connected to the $\Lambda(1670)$ at the physical point, allowing for the investigation of this state without the need for the three-body unitarity. 
\section{Conclusion and outlook}

We have calculated the energy levels of the singlet and octet irreducible representations of the meson-baryon interaction at the $\mathrm{SU}(3)$-flavor point. We have used meson-baryon operators that belong to the different irreps of that symmetry group. We have found that the ground states are strongly attractive with large negative energy shifts relative to the threshold, suggesting they may be bound states.

The energy associated with the singlet state is the lowest and significantly distinct from the octets. In the octet levels we found two distinct energy levels at one sigma. However, they are compatible with being degenerate at two sigmas, so we cannot conclusively say that they are not degenerate. 

We have compared our results with the ones predicted from the UCHPT analysis of \cite{Pittler:2025upn}. We find that the fit $F_{12}$, which describes all available lattice (BaSC $M_\pi\approx 200\MeV$) and experimental data only agrees with one of the octets and misses the other two energy levels entirely. This shows that our data, indeed, provides new quantitative constraints on the scattering amplitudes. On the other hand, the fit $F_{16}$ of that reference predicts qualitatively similar finite-volume spectrum as the lattice QCD determined one. The octet states agree within two sigmas, while the singlet is slightly further away. Analytically continuing this solution to the complex energies we find that, indeed all three energy eigenvalues correspond to a bound state (pole on the real axis below the meson-baryon threshold.) confirming our previous presumption. We have then connected all three QCD configurations (physical, BaSC, and $\mathrm{SU}(3)$-flavor point) through a trajectory. Following this trajectory, we have uniquely identified the movement of the poles. We see that the $\Lambda(1380)$ pole originates from the singlet, while the $\Lambda(1405)$ originates from the lowest bound octet. Encouraged by this, we have performed a re-fit, $F_{16'}$, now also including our new results and confirming this connection between the resonance poles at the physical and bound states at the heavy mass $\mathrm{SU}(3)$-flavor symmetric point. 

An important step in the future would be to have some control over the dependence of the pion mass. For this reason, we aim to extend this calculation to a set of gauge configurations with a pion mass of approximately 450 MeV, which in turn will allow for more inputs to constrain CHPT. Furthermore, the detailed global analysis of all available data including systematics associated with scale setting and other intricacies is planned.

\begin{acknowledgments}
We would like to thank A. Walker-Loud for providing us with the gauge configurations used in this work.
This work was partly funded by the Deutsche Forschungsgemeinschaft (DFG, German Research Foundation) as part of the CRC 1639 NuMeriQS – Project number 511713970. This work has been supported by the MKW NRW under the funding code NW21-024-A as part of NRW-FAIR.
The work of MM was further funded through the Heisenberg Programme by the Deutsche Forschungsgemeinschaft (DFG, German Research Foundation) – 532635001. 
The work of FP was supported by the projects PulseQCD,
DeNuTra, MuonHVP(EXCELLENCE/0524/0269, EXCELLENCE/524/0455, EXCELLENCE/524/0017) cofinanced by the 
European Regional Development Fund and the Republic of Cyprus through the Research and
Innovation Foundation, and under Germany’s Excellence Strategy –
  EXC 3107 – Project-ID 533766364 in the Color-meets-flavor cluster of
  excellence. HY acknowledges support from NSFC under Grant No.~124B2096. The open source
software packages R~\cite{R-base} and hadron~\cite{hadron} have been used. The authors gratefully acknowledge the access to the Marvin cluster and the HPC@HRZ team of the University of Bonn .
\end{acknowledgments}

\bibliography{bibliography}

@article{Dalitz:1959dq,
    author = "Dalitz, R. H. and Tuan, S. F.",
    title = "{The energy dependence of low energy K- -proton processes}",
    doi = "10.1016/0003-4916(59)90064-8",
    journal = "Annals Phys.",
    volume = "8",
    pages = "100--118",
    year = "1959"
}

@article{Pittler:2025upn,
    author = "Pittler, Ferenc and Mai, Maxim and Mei{\ss}ner, Ulf-G. and Ferguson, Ryan F. and Hurck, Peter and Ireland, David G. and McKinnon, Bryan",
    title = "{Universal parameters of the {\ensuremath{\Lambda}}(1380), the {\ensuremath{\Lambda}}(1405), and their isospin partners from a combined analysis of lattice QCD and experimental results}",
    eprint = "2507.14283",
    archivePrefix = "arXiv",
    primaryClass = "hep-ph",
    doi = "10.1103/ls4c-6f2y",
    journal = "Phys. Rev. D",
    volume = "112",
    number = "7",
    pages = "074037",
    year = "2025"
}

@article{Mai:2020ltx,
    author = "Mai, Maxim",
    title = "{Review of the ${\Lambda }$(1405) A curious case of a strangeness resonance}",
    eprint = "2010.00056",
    archivePrefix = "arXiv",
    primaryClass = "nucl-th",
    doi = "10.1140/epjs/s11734-021-00144-7",
    journal = "Eur. Phys. J. ST",
    volume = "230",
    number = "6",
    pages = "1593--1607",
    year = "2021"
}

@article{Meissner:2020khl,
    author = "Mei{\ss}ner, Ulf-G.",
    title = "{Two-pole structures in QCD: Facts, not fantasy!}",
    eprint = "2005.06909",
    archivePrefix = "arXiv",
    primaryClass = "hep-ph",
    doi = "10.3390/sym12060981",
    journal = "Symmetry",
    volume = "12",
    number = "6",
    pages = "981",
    year = "2020"
}

@article{Hyodo:2020czb,
    author = "Hyodo, Tetsuo and Niiyama, Masayuki",
    title = "{QCD and the strange baryon spectrum}",
    eprint = "2010.07592",
    archivePrefix = "arXiv",
    primaryClass = "hep-ph",
    doi = "10.1016/j.ppnp.2021.103868",
    journal = "Prog. Part. Nucl. Phys.",
    volume = "120",
    pages = "103868",
    year = "2021"
}

@article{Niiyama:2009zza,
    author = "Niiyama, Masayuki",
    editor = "Tserruya, Itzhak and Gal, Avraham and Ashery, Daniel",
    collaboration = "LEPS TPC",
    title = "{Photoproduction of Sigma**0(1385) and Lambda (1405) on the proton near threshold}",
    doi = "10.1016/j.nuclphysa.2009.05.051",
    journal = "Nucl. Phys. A",
    volume = "827",
    pages = "261C--263C",
    year = "2009"
}

@article{CLAS:2013rjt,
    author = "Moriya, K. and others",
    collaboration = "CLAS",
    title = "{Measurement of the {\ensuremath{\Sigma}}{\ensuremath{\pi}} photoproduction line shapes near the {\ensuremath{\Lambda}}(1405)}",
    eprint = "1301.5000",
    archivePrefix = "arXiv",
    primaryClass = "nucl-ex",
    reportNumber = "JLAB-PHY-13-1690",
    doi = "10.1103/PhysRevC.87.035206",
    journal = "Phys. Rev. C",
    volume = "87",
    number = "3",
    pages = "035206",
    year = "2013"
}

@article{CLAS:2013rxx,
    author = "Moriya, K. and others",
    collaboration = "CLAS",
    title = "{Differential Photoproduction Cross Sections of the $\Sigma^0(1385)$, $\Lambda(1405)$, and $\Lambda(1520)$}",
    eprint = "1305.6776",
    archivePrefix = "arXiv",
    primaryClass = "nucl-ex",
    reportNumber = "JLAB-PHY-13-1744",
    doi = "10.1103/PhysRevC.88.045201",
    journal = "Phys. Rev. C",
    volume = "88",
    pages = "045201",
    year = "2013",
    note = "[Addendum: Phys.Rev.C 88, 049902 (2013)]"
}

@article{CLAS:2014tbc,
    author = "Moriya, K. and others",
    collaboration = "CLAS",
    title = "{Spin and parity measurement of the Lambda(1405) baryon}",
    eprint = "1402.2296",
    archivePrefix = "arXiv",
    primaryClass = "hep-ex",
    reportNumber = "JLAB-PHY-14-1848",
    doi = "10.1103/PhysRevLett.112.082004",
    journal = "Phys. Rev. Lett.",
    volume = "112",
    number = "8",
    pages = "082004",
    year = "2014"
}

@article{BGOOD:2021sog,
    author = "Scheluchin, G. and others",
    collaboration = "BGOOD",
    title = "{Photoproduction of K+{\ensuremath{\Lambda}}(1405){\textrightarrow}K+{\ensuremath{\pi}}0{\ensuremath{\Sigma}}0 extending to forward angles and low momentum transfer}",
    eprint = "2108.12235",
    archivePrefix = "arXiv",
    primaryClass = "nucl-ex",
    doi = "10.1016/j.physletb.2022.137375",
    journal = "Phys. Lett. B",
    volume = "833",
    pages = "137375",
    year = "2022"
}

@article{J-PARCE31:2022plu,
    author = "Aikawa, S. and others",
    collaboration = "J-PARC E31",
    title = "{Pole position of {\ensuremath{\Lambda}}(1405) measured in d(K{\ensuremath{-}},n){\ensuremath{\pi}}{\ensuremath{\Sigma}} reactions}",
    eprint = "2209.08254",
    archivePrefix = "arXiv",
    primaryClass = "nucl-ex",
    doi = "10.1016/j.physletb.2022.137637",
    journal = "Phys. Lett. B",
    volume = "837",
    pages = "137637",
    year = "2023"
}

@article{Anisovich:2020lec,
    author = "Anisovich, A. V. and Sarantsev, A. V. and Nikonov, V. A. and Burkert, V. and Schumacher, R. A. and Thoma, U. and Klempt, E.",
    title = "{Hyperon III: $K^{-}p - \pi \Sigma $ coupled-channel dynamics in the $\Lambda (1405)$ mass region}",
    doi = "10.1140/epja/s10050-020-00142-8",
    journal = "Eur. Phys. J. A",
    volume = "56",
    number = "5",
    pages = "139",
    year = "2020"
}

@article{Roca:2015tea,
    author = "Roca, Luis and Mai, Maxim and Oset, Eulogio and Mei{\ss}ner, Ulf-G.",
    title = "{Predictions for the ${{\Lambda }_b \rightarrow J/\psi ~ \Lambda (1405)}$ decay}",
    eprint = "1503.02936",
    archivePrefix = "arXiv",
    primaryClass = "hep-ph",
    doi = "10.1140/epjc/s10052-015-3438-1",
    journal = "Eur. Phys. J. C",
    volume = "75",
    number = "5",
    pages = "218",
    year = "2015"
}

@article{Mai:2014xna,
    author = "Mai, Maxim and Mei{\ss}ner, Ulf-G.",
    title = "{Constraints on the chiral unitary $\bar KN$ amplitude from $\pi\Sigma K^+$ photoproduction data}",
    eprint = "1411.7884",
    archivePrefix = "arXiv",
    primaryClass = "hep-ph",
    doi = "10.1140/epja/i2015-15030-3",
    journal = "Eur. Phys. J. A",
    volume = "51",
    number = "3",
    pages = "30",
    year = "2015"
}

@article{Lutz:2004sg,
    author = "Lutz, Matthias F. M. and Soyeur, Madeleine",
    title = "{The Associated photoproduction of positive kaons and pi0 Lambda or pi Sigma pairs in the region of the Sigma(1385) and Lambda(1405) resonances}",
    eprint = "nucl-th/0407115",
    archivePrefix = "arXiv",
    reportNumber = "DAPNIA-04-201",
    doi = "10.1016/j.nuclphysa.2004.11.007",
    journal = "Nucl. Phys. A",
    volume = "748",
    pages = "499--512",
    year = "2005"
}

@article{SIDDHARTA:2011dsy,
    author = "Bazzi, M. and others",
    collaboration = "SIDDHARTA",
    title = "{A New Measurement of Kaonic Hydrogen X-rays}",
    eprint = "1105.3090",
    archivePrefix = "arXiv",
    primaryClass = "nucl-ex",
    doi = "10.1016/j.physletb.2011.09.011",
    journal = "Phys. Lett. B",
    volume = "704",
    pages = "113--117",
    year = "2011"
}

@article{Piscicchia:2018rez,
    author = "Piscicchia, K. and others",
    title = "{First measurement of the $K^- n \to \Lambda \pi^-$ non-resonant transition amplitude below threshold}",
    doi = "10.1016/j.physletb.2018.05.025",
    journal = "Phys. Lett. B",
    volume = "782",
    pages = "339--345",
    year = "2018"
}

@article{Piscicchia:2022wmd,
    author = "Piscicchia, Kristian and others",
    title = "{First simultaneous K{\ensuremath{-}}p{\textrightarrow}{\ensuremath{\Sigma}}0{\ensuremath{\pi}}0,~{\ensuremath{\Lambda}}{\ensuremath{\pi}}0 cross section~measurements at 98 MeV/c}",
    eprint = "2210.10342",
    archivePrefix = "arXiv",
    primaryClass = "nucl-ex",
    doi = "10.1103/PhysRevC.108.055201",
    journal = "Phys. Rev. C",
    volume = "108",
    number = "5",
    pages = "055201",
    year = "2023"
}

@article{Curceanu:2026zjg,
    author = "Curceanu, Catalina and Sgaramella, Francesco and Bazzi, Massimiliano and Hashimoto, Tadashi and Iliescu, Mihail and Scordo, Alessandro and Sirghi, Diana and Sirghi, Florin",
    title = "{Light kaonic atoms as probes of fundamental interactions in strange systems}",
    doi = "10.1016/j.ppnp.2026.104226",
    journal = "Prog. Part. Nucl. Phys.",
    volume = "147",
    pages = "104226",
    year = "2026"
}

@article{Jido:2003cb,
    author = "Jido, D. and Oller, J. A. and Oset, E. and Ramos, A. and Meissner, U. G.",
    title = "{Chiral dynamics of the two Lambda(1405) states}",
    eprint = "nucl-th/0303062",
    archivePrefix = "arXiv",
    doi = "10.1016/S0375-9474(03)01598-7",
    journal = "Nucl. Phys. A",
    volume = "725",
    pages = "181--200",
    year = "2003"
}

@article{Bruns:2021krp,
    author = "Bruns, P. C. and Ciepl{\'y}, A.",
    title = "{SU(3) flavor symmetry considerations for the K{\textasciimacron}N coupled channels system}",
    eprint = "2109.03109",
    archivePrefix = "arXiv",
    primaryClass = "hep-ph",
    doi = "10.1016/j.nuclphysa.2021.122378",
    journal = "Nucl. Phys. A",
    volume = "1019",
    pages = "122378",
    year = "2022"
}

@article{Guo:2023wes,
    author = "Guo, Feng-Kun and Kamiya, Yuki and Mai, Maxim and Mei{\ss}ner, Ulf-G.",
    title = "{New insights into the nature of the {\ensuremath{\Lambda}}(1380) and {\ensuremath{\Lambda}}(1405) resonances away from the SU(3) limit}",
    eprint = "2308.07658",
    archivePrefix = "arXiv",
    primaryClass = "hep-ph",
    doi = "10.1016/j.physletb.2023.138264",
    journal = "Phys. Lett. B",
    volume = "846",
    pages = "138264",
    year = "2023"
}

@article{Mai:2022eur,
    author = "Mai, Maxim and Mei{\ss}ner, Ulf-G. and Urbach, Carsten",
    title = "{Towards a theory of hadron resonances}",
    eprint = "2206.01477",
    archivePrefix = "arXiv",
    primaryClass = "hep-ph",
    doi = "10.1016/j.physrep.2022.11.005",
    journal = "Phys. Rept.",
    volume = "1001",
    pages = "1--66",
    year = "2023"
}

@article{Torok:2009dg,
    author = "Torok, Aaron and Beane, Silas R. and Detmold, William and Luu, Thomas C. and Orginos, Kostas and Parreno, Assumpta and Savage, Martin J. and Walker-Loud, Andre",
    title = "{Meson-Baryon Scattering Lengths from Mixed-Action Lattice QCD}",
    eprint = "0907.1913",
    archivePrefix = "arXiv",
    primaryClass = "hep-lat",
    reportNumber = "UNH-09-03, JLAB-THY-09-1021, ICCUB-09-217, ATHENA-PUB-09-017, NT-UW-09-16",
    doi = "10.1103/PhysRevD.81.074506",
    journal = "Phys. Rev. D",
    volume = "81",
    pages = "074506",
    year = "2010"
}

@article{Gubler:2016viv,
    author = "Gubler, Philipp and Takahashi, Toru T. and Oka, Makoto",
    title = "{Flavor structure of $\Lambda$ baryons from lattice QCD: From strange to charm quarks}",
    eprint = "1609.01889",
    archivePrefix = "arXiv",
    primaryClass = "hep-lat",
    doi = "10.1103/PhysRevD.94.114518",
    journal = "Phys. Rev. D",
    volume = "94",
    number = "11",
    pages = "114518",
    year = "2016"
}

@article{Menadue:2011pd,
    author = "Menadue, Benjamin J. and Kamleh, Waseem and Leinweber, Derek B. and Mahbub, M. Selim",
    title = "{Isolating the $\Lambda(1405)$ in Lattice QCD}",
    eprint = "1109.6716",
    archivePrefix = "arXiv",
    primaryClass = "hep-lat",
    reportNumber = "ADP-11-28-T750",
    doi = "10.1103/PhysRevLett.108.112001",
    journal = "Phys. Rev. Lett.",
    volume = "108",
    pages = "112001",
    year = "2012"
}

@article{Engel:2012qp,
    author = {Engel, Georg P. and Lang, C. B. and Sch{\"a}fer, Andreas},
    collaboration = "BGR (Bern-Graz-Regensburg)",
    title = "{Low-lying $\Lambda$ baryons from the lattice}",
    eprint = "1212.2032",
    archivePrefix = "arXiv",
    primaryClass = "hep-lat",
    doi = "10.1103/PhysRevD.87.034502",
    journal = "Phys. Rev. D",
    volume = "87",
    number = "3",
    pages = "034502",
    year = "2013"
}

@article{Engel:2013ig,
    author = {Engel, Georg P. and Lang, C. B. and Mohler, Daniel and Sch{\"a}fer, Andreas},
    collaboration = "BGR",
    title = "{QCD with Two Light Dynamical Chirally Improved Quarks: Baryons}",
    eprint = "1301.4318",
    archivePrefix = "arXiv",
    primaryClass = "hep-lat",
    reportNumber = "FERMILAB-PUB-13-028-T",
    doi = "10.1103/PhysRevD.87.074504",
    journal = "Phys. Rev. D",
    volume = "87",
    number = "7",
    pages = "074504",
    year = "2013"
}

@article{Nemoto:2003ft,
    author = "Nemoto, Y. and Nakajima, N. and Matsufuru, H. and Suganuma, H.",
    title = "{Negative parity baryons in quenched anisotropic lattice QCD}",
    eprint = "hep-lat/0302013",
    archivePrefix = "arXiv",
    doi = "10.1103/PhysRevD.68.094505",
    journal = "Phys. Rev. D",
    volume = "68",
    pages = "094505",
    year = "2003"
}

@article{Burch:2006cc,
    author = "Burch, Tommy and Gattringer, Christof and Glozman, Leonid Ya. and Hagen, Christian and Hierl, Dieter and Lang, C. B. and Schafer, Andreas",
    title = "{Excited hadrons on the lattice: Baryons}",
    eprint = "hep-lat/0604019",
    archivePrefix = "arXiv",
    doi = "10.1103/PhysRevD.74.014504",
    journal = "Phys. Rev. D",
    volume = "74",
    pages = "014504",
    year = "2006"
}

@article{Takahashi:2009bu,
    author = "Takahashi, Toru T. and Oka, Makoto",
    title = "{Low-lying Lambda Baryons with spin 1/2 in Two-flavor Lattice QCD}",
    eprint = "0910.0686",
    archivePrefix = "arXiv",
    primaryClass = "hep-lat",
    doi = "10.1103/PhysRevD.81.034505",
    journal = "Phys. Rev. D",
    volume = "81",
    pages = "034505",
    year = "2010"
}

@article{Meinel:2021grq,
    author = "Meinel, Stefan and Rendon, Gumaro",
    title = "{Charm-baryon semileptonic decays and the strange {\ensuremath{\Lambda}}* resonances: New insights from lattice QCD}",
    eprint = "2107.13084",
    archivePrefix = "arXiv",
    primaryClass = "hep-ph",
    doi = "10.1103/PhysRevD.105.L051505",
    journal = "Phys. Rev. D",
    volume = "105",
    number = "5",
    pages = "L051505",
    year = "2022"
}

@article{Hall:2014uca,
    author = "Hall, Jonathan M. M. and Kamleh, Waseem and Leinweber, Derek B. and Menadue, Benjamin J. and Owen, Benjamin J. and Thomas, Anthony W. and Young, Ross D.",
    title = "{Lattice QCD Evidence that the {\ensuremath{\Lambda}}(1405) Resonance is an Antikaon-Nucleon Molecule}",
    eprint = "1411.3402",
    archivePrefix = "arXiv",
    primaryClass = "hep-lat",
    reportNumber = "ADP-14-34-T893",
    doi = "10.1103/PhysRevLett.114.132002",
    journal = "Phys. Rev. Lett.",
    volume = "114",
    number = "13",
    pages = "132002",
    year = "2015"
}

@article{BaryonScatteringBaSc:2023zvt,
    author = "Bulava, John and others",
    collaboration = "Baryon Scattering (BaSc)",
    title = "{Two-Pole Nature of the {\ensuremath{\Lambda}}(1405) resonance from Lattice QCD}",
    eprint = "2307.10413",
    archivePrefix = "arXiv",
    primaryClass = "hep-lat",
    reportNumber = "MIT-CTP/5579",
    doi = "10.1103/PhysRevLett.132.051901",
    journal = "Phys. Rev. Lett.",
    volume = "132",
    number = "5",
    pages = "051901",
    year = "2024"
}

@article{BaryonScatteringBaSc:2023ori,
    author = "Bulava, John and others",
    collaboration = "Baryon Scattering (BaSc)",
    title = "{Lattice QCD study of {\ensuremath{\pi}}{\ensuremath{\Sigma}}-K{\textasciimacron}N scattering and the {\ensuremath{\Lambda}}(1405) resonance}",
    eprint = "2307.13471",
    archivePrefix = "arXiv",
    primaryClass = "hep-lat",
    reportNumber = "MIT-CTP/5580",
    doi = "10.1103/PhysRevD.109.014511",
    journal = "Phys. Rev. D",
    volume = "109",
    number = "1",
    pages = "014511",
    year = "2024"
}

@article{Murakami:2023phq,
    author = "Murakami, Kotaro and Aoki, Sinya",
    title = "{Study on Lambda(1405) in the flavor SU(3) limit in the HAL QCD method}",
    eprint = "2311.17421",
    archivePrefix = "arXiv",
    primaryClass = "hep-lat",
    reportNumber = "YITP-23-153, RIKEN-iTHEMS-Report-23",
    doi = "10.22323/1.453.0063",
    journal = "PoS",
    volume = "LATTICE2023",
    pages = "063",
    year = "2024"
}

@article{Murakami:2025oig,
    author = "Murakami, Kotaro and Aoki, Sinya",
    title = "{$\Lambda$(1405) in the flavor SU(3) limit using a separable potential in the HAL QCD method}",
    eprint = "2501.17423",
    archivePrefix = "arXiv",
    primaryClass = "hep-lat",
    reportNumber = "RIKEN-iTHEMS-Report-25, YITP-25-11",
    doi = "10.22323/1.466.0101",
    journal = "PoS",
    volume = "LATTICE2024",
    pages = "101",
    year = "2025"
}

@article{Lutz:2024ubv,
    author = "Lutz, Matthias F. M. and Heo, Yonggoo and Hudspith, Renwick J.",
    title = "{QCD in the chiral SU(3) limit from baryon masses on lattice QCD ensembles}",
    eprint = "2406.07442",
    archivePrefix = "arXiv",
    primaryClass = "hep-lat",
    doi = "10.1103/PhysRevD.110.094046",
    journal = "Phys. Rev. D",
    volume = "110",
    number = "9",
    pages = "094046",
    year = "2024"
}

@article{Ren:2024frr,
    author = "Ren, Xiu-Lei",
    title = "{Light-quark mass dependence of the {\ensuremath{\Lambda}}(1405) resonance}",
    eprint = "2404.02720",
    archivePrefix = "arXiv",
    primaryClass = "hep-ph",
    doi = "10.1016/j.physletb.2024.138802",
    journal = "Phys. Lett. B",
    volume = "855",
    pages = "138802",
    year = "2024"
}

@book{Georgi:1999wka,
    author = "Georgi, Howard",
    title = "{Lie algebras in particle physics}",
    edition = "2nd ed.",
    publisher = "Perseus Books",
    address = "Reading, MA",
    volume = "54",
    year = "1999"
}

@article{Gregory:2021rgy,
    author = "Gregory, Eric B. and Guo, Feng-Kun and Hanhart, Christoph and Krieg, Stefan and Luu, Thomas",
    title = "{Confirmation of the existence of an exotic state in the $\pi D$ system}",
    eprint = "2106.15391",
    archivePrefix = "arXiv",
    primaryClass = "hep-ph",
    month = "6",
    year = "2021"
}

@article{Gregory:2025ium,
    author = "Gregory, Eric B. and Guo, Feng-Kun and Hanhart, Christoph and Krieg, Stefan and Luu, Thomas",
    title = "{Exclusion of a diquark{\textendash}anti-diquark structure for the lightest positive-parity charmed mesons}",
    eprint = "2503.23954",
    archivePrefix = "arXiv",
    primaryClass = "hep-lat",
    doi = "10.1140/epja/s10050-025-01665-8",
    journal = "Eur. Phys. J. A",
    volume = "61",
    number = "10",
    pages = "226",
    year = "2025"
}

@article{Horz:2020zvv,
    author = {H{\"o}rz, Ben and others},
    title = "{Two-nucleon S-wave interactions at the $SU(3)$ flavor-symmetric point with $m_{ud}\simeq m_s^{\rm phys}$: A first lattice QCD calculation with the stochastic Laplacian Heaviside method}",
    eprint = "2009.11825",
    archivePrefix = "arXiv",
    primaryClass = "hep-lat",
    reportNumber = "LLNL-JRNL-813871, RIKEN-iTHEMS-Report-20, MITP/20-055",
    doi = "10.1103/PhysRevC.103.014003",
    journal = "Phys. Rev. C",
    volume = "103",
    number = "1",
    pages = "014003",
    year = "2021"
}

@article{BaryonScattering:2025ziz,
    author = "Bulava, John and others",
    collaboration = "Baryon Scattering",
    title = "{Di-nucleons do not form bound states at heavy pion mass}",
    eprint = "2505.05547",
    archivePrefix = "arXiv",
    primaryClass = "hep-lat",
    reportNumber = "LLNL-JRNL-2005660",
    doi = "10.1103/d2hg-h6d4",
    journal = "Phys. Rev. C",
    volume = "113",
    number = "2",
    pages = "024002",
    year = "2026"
}

@article{HadronSpectrum:2009krc,
    author = "Peardon, Michael and Bulava, John and Foley, Justin and Morningstar, Colin and Dudek, Jozef and Edwards, Robert G. and Joo, Balint and Lin, Huey-Wen and Richards, David G. and Juge, Keisuke Jimmy",
    collaboration = "Hadron Spectrum",
    title = "{A Novel quark-field creation operator construction for hadronic physics in lattice QCD}",
    eprint = "0905.2160",
    archivePrefix = "arXiv",
    primaryClass = "hep-lat",
    reportNumber = "JLAB-THY-09-985",
    doi = "10.1103/PhysRevD.80.054506",
    journal = "Phys. Rev. D",
    volume = "80",
    pages = "054506",
    year = "2009"
}

@article{Bruno:2016plf,
    author = "Bruno, Mattia and Korzec, Tomasz and Schaefer, Stefan",
    title = "{Setting the scale for the CLS $2 + 1$ flavor ensembles}",
    eprint = "1608.08900",
    archivePrefix = "arXiv",
    primaryClass = "hep-lat",
    reportNumber = "DESY-16-162, WUB-16-05",
    doi = "10.1103/PhysRevD.95.074504",
    journal = "Phys. Rev. D",
    volume = "95",
    number = "7",
    pages = "074504",
    year = "2017"
}

@article{DallaBrida:2018tpn,
    author = "Dalla Brida, Mattia and Korzec, Tomasz and Sint, Stefan and Vilaseca, Pol",
    title = "{High precision renormalization of the flavour non-singlet Noether currents in lattice QCD with Wilson quarks}",
    eprint = "1808.09236",
    archivePrefix = "arXiv",
    primaryClass = "hep-lat",
    doi = "10.1140/epjc/s10052-018-6514-5",
    journal = "Eur. Phys. J. C",
    volume = "79",
    number = "1",
    pages = "23",
    year = "2019"
}

@article{Kuberski:2024pms,
    author = {Kuberski, Simon and Joswig, Fabian and Collins, Sara and Heitger, Jochen and S{\"o}ldner, Wolfgang},
    collaboration = "RQCD, ALPHA",
    title = "{D and D$_{s}$ decay constants in N$_{f}$ = 2 + 1 QCD with Wilson fermions}",
    eprint = "2405.04506",
    archivePrefix = "arXiv",
    primaryClass = "hep-lat",
    reportNumber = "CERN-TH-2024-052, MITP-24-047, MS-TP-24-09",
    doi = "10.1007/JHEP07(2024)090",
    journal = "JHEP",
    volume = "07",
    pages = "090",
    year = "2024"
}

@article{Jiang:2024lto,
    author = "Jiang, Xiangyu and Shi, Chunjiang and Chen, Ying and Gong, Ming and Yang, Yi-Bo",
    title = "{Use QUDA for lattice QCD calculation with Python}",
    eprint = "2411.08461",
    archivePrefix = "arXiv",
    primaryClass = "hep-lat",
    month = "11",
    year = "2024"
}

@article{Edwards:2004sx,
    author = "Edwards, Robert G. and Joo, Balint",
    editor = "Bodwin, Geoffrey T. and Sinclair, D. K. and Eichten, E. and Holmgren, D. and Kronfeld, Andreas S. and Mackenzie, P. and Okamoto, M. and Simone, J. and El-Khadra, Aida X.",
    collaboration = "SciDAC, LHPC, UKQCD",
    title = "{The Chroma software system for lattice QCD}",
    eprint = "hep-lat/0409003",
    archivePrefix = "arXiv",
    reportNumber = "JLAB-THY-04-54",
    doi = "10.1016/j.nuclphysbps.2004.11.254",
    journal = "Nucl. Phys. B Proc. Suppl.",
    volume = "140",
    pages = "832",
    year = "2005"
}

@article{Clark:2009wm,
    author = "Clark, M. A. and Babich, R. and Barros, K. and Brower, R. C. and Rebbi, C.",
    collaboration = "QUDA",
    title = "{Solving Lattice QCD systems of equations using mixed precision solvers on GPUs}",
    eprint = "0911.3191",
    archivePrefix = "arXiv",
    primaryClass = "hep-lat",
    doi = "10.1016/j.cpc.2010.05.002",
    journal = "Comput. Phys. Commun.",
    volume = "181",
    pages = "1517--1528",
    year = "2010"
}

@article{Michael:1982gb,
    author = "Michael, Christopher and Teasdale, I.",
    title = "{Extracting Glueball Masses From Lattice {QCD}}",
    reportNumber = "LTH 96",
    doi = "10.1016/0550-3213(83)90674-0",
    journal = "Nucl. Phys. B",
    volume = "215",
    pages = "433--446",
    year = "1983"
}

@article{Luscher:1990ck,
    author = "Luscher, Martin and Wolff, Ulli",
    title = "{How to Calculate the Elastic Scattering Matrix in Two-dimensional Quantum Field Theories by Numerical Simulation}",
    reportNumber = "DESY-90-010",
    doi = "10.1016/0550-3213(90)90540-T",
    journal = "Nucl. Phys. B",
    volume = "339",
    pages = "222--252",
    year = "1990"
}

@article{Blossier:2009kd,
    author = "Blossier, Benoit and Della Morte, Michele and von Hippel, Georg and Mendes, Tereza and Sommer, Rainer",
    title = "{On the generalized eigenvalue method for energies and matrix elements in lattice field theory}",
    eprint = "0902.1265",
    archivePrefix = "arXiv",
    primaryClass = "hep-lat",
    reportNumber = "DESY-09-014, SFB-CPP-09-10, MKPH-T-09-01, LPT-ORSAY-09-05",
    doi = "10.1088/1126-6708/2009/04/094",
    journal = "JHEP",
    volume = "04",
    pages = "094",
    year = "2009"
}

@article{Luscher:1986pf,
    author = "Luscher, M.",
    title = "{Volume Dependence of the Energy Spectrum in Massive Quantum Field Theories. 2. Scattering States}",
    reportNumber = "DESY-86-034",
    doi = "10.1007/BF01211097",
    journal = "Commun. Math. Phys.",
    volume = "105",
    pages = "153--188",
    year = "1986"
}

@article{Luscher:1990ux,
    author = "Luscher, Martin",
    title = "{Two particle states on a torus and their relation to the scattering matrix}",
    reportNumber = "DESY-90-131",
    doi = "10.1016/0550-3213(91)90366-6",
    journal = "Nucl. Phys. B",
    volume = "354",
    pages = "531--578",
    year = "1991"
}

@article{Luu:2011ep,
    author = "Luu, Thomas and Savage, Martin J.",
    title = "{Extracting Scattering Phase-Shifts in Higher Partial-Waves from Lattice QCD Calculations}",
    eprint = "1101.3347",
    archivePrefix = "arXiv",
    primaryClass = "hep-lat",
    doi = "10.1103/PhysRevD.83.114508",
    journal = "Phys. Rev. D",
    volume = "83",
    pages = "114508",
    year = "2011"
}

@article{Gasser:1987rb,
    author = "Gasser, J. and Sainio, M. E. and Svarc, A.",
    title = "{Nucleons with chiral loops}",
    reportNumber = "BUTP-87-17",
    doi = "10.1016/0550-3213(88)90108-3",
    journal = "Nucl. Phys. B",
    volume = "307",
    pages = "779--853",
    year = "1988"
}

@article{Bernard:1992qa,
    author = "Bernard, Veronique and Kaiser, Norbert and Kambor, Joachim and Meissner, Ulf G.",
    title = "{Chiral structure of the nucleon}",
    reportNumber = "BUTP-92-15, CRN-92-24, TUM-T31-28-92",
    doi = "10.1016/0550-3213(92)90615-I",
    journal = "Nucl. Phys. B",
    volume = "388",
    pages = "315--345",
    year = "1992"
}

@article{Tang:1996ca,
    author = "Tang, Hua-Bin",
    title = "{A New approach to chiral perturbation theory for matter fields}",
    eprint = "hep-ph/9607436",
    archivePrefix = "arXiv",
    reportNumber = "NUC-MINN-96-11-T",
    month = "7",
    year = "1996"
}

@article{Ellis:1997kc,
    author = "Ellis, Paul J. and Tang, Hua-Bin",
    title = "{Pion nucleon scattering in a new approach to chiral perturbation theory}",
    eprint = "hep-ph/9709354",
    archivePrefix = "arXiv",
    reportNumber = "NUC-MINN-97-8-T",
    doi = "10.1103/PhysRevC.57.3356",
    journal = "Phys. Rev. C",
    volume = "57",
    pages = "3356--3375",
    year = "1998"
}

@article{Frink:2006hx,
    author = "Frink, Matthias and Meissner, Ulf-G.",
    title = "{On the chiral effective meson-baryon Lagrangian at third order}",
    eprint = "hep-ph/0609256",
    archivePrefix = "arXiv",
    reportNumber = "HISKP-TH-06-29",
    doi = "10.1140/epja/i2006-10105-x",
    journal = "Eur. Phys. J. A",
    volume = "29",
    pages = "255--260",
    year = "2006"
}

@article{Holmberg:2018dtv,
    author = "Holmberg, M{\r{a}}ns and Leupold, Stefan",
    title = "{The relativistic chiral Lagrangian for decuplet and octet baryons at next-to-leading order}",
    eprint = "1802.05168",
    archivePrefix = "arXiv",
    primaryClass = "hep-ph",
    doi = "10.1140/epja/i2018-12533-3",
    journal = "Eur. Phys. J. A",
    volume = "54",
    number = "6",
    pages = "103",
    year = "2018"
}

@article{Song:2024fae,
    author = "Song, Chuan-Qiang and Sun, Hao and Yu, Jiang-Hao",
    title = "{Complete CP-eigen bases of meson-baryon chiral lagrangian up to p$^{5}$-order}",
    eprint = "2404.15047",
    archivePrefix = "arXiv",
    primaryClass = "hep-ph",
    doi = "10.1007/JHEP09(2024)171",
    journal = "JHEP",
    volume = "09",
    pages = "171",
    year = "2024"
}

@article{Lutz:2023xpi,
    author = "Lutz, Matthias F. M. and Heo, Yonggoo and Guo, Xiao-Yu",
    title = "{Low-energy constants in the chiral Lagrangian with baryon octet and decuplet fields from Lattice QCD data on CLS ensembles}",
    eprint = "2301.06837",
    archivePrefix = "arXiv",
    primaryClass = "hep-lat",
    doi = "10.1140/epjc/s10052-023-11556-1",
    journal = "Eur. Phys. J. C",
    volume = "83",
    number = "5",
    pages = "440",
    year = "2023"
}

@article{Mai:2009ce,
    author = "Mai, Maxim and Bruns, Peter C. and Kubis, Bastian and Meissner, Ulf-G.",
    title = "{Aspects of meson-baryon scattering in three and two-flavor chiral perturbation theory}",
    eprint = "0905.2810",
    archivePrefix = "arXiv",
    primaryClass = "hep-ph",
    reportNumber = "HISKP-TH-09-12, FZJ-IKP-TH-2009-10",
    doi = "10.1103/PhysRevD.80.094006",
    journal = "Phys. Rev. D",
    volume = "80",
    pages = "094006",
    year = "2009"
}

@article{Oller:2000fj,
    author = "Oller, J. A. and Meissner, Ulf G.",
    title = "{Chiral dynamics in the presence of bound states: Kaon nucleon interactions revisited}",
    eprint = "hep-ph/0011146",
    archivePrefix = "arXiv",
    reportNumber = "FZJ-IKP-TH-2000-26",
    doi = "10.1016/S0370-2693(01)00078-8",
    journal = "Phys. Lett. B",
    volume = "500",
    pages = "263--272",
    year = "2001"
}

@article{Lutz:2001yb,
    author = "Lutz, M. F. M. and Kolomeitsev, E. E.",
    title = "{Relativistic chiral SU(3) symmetry, large N(c) sum rules and meson baryon scattering}",
    eprint = "nucl-th/0105042",
    archivePrefix = "arXiv",
    reportNumber = "GSI-PREPRINT-2001-12, ECT-2001-10",
    doi = "10.1016/S0375-9474(01)01312-4",
    journal = "Nucl. Phys. A",
    volume = "700",
    pages = "193--308",
    year = "2002"
}

@article{Borasoy:2005ie,
    author = "Borasoy, B. and Nissler, R. and Weise, W.",
    title = "{Chiral dynamics of kaon-nucleon interactions, revisited}",
    eprint = "hep-ph/0505239",
    archivePrefix = "arXiv",
    doi = "10.1140/epja/i2005-10079-1",
    journal = "Eur. Phys. J. A",
    volume = "25",
    pages = "79--96",
    year = "2005"
}

@article{Oller:2005ig,
    author = "Oller, Jose A. and Prades, Joaquim and Verbeni, Michela",
    title = "{Surprises in threshold antikaon-nucleon physics}",
    eprint = "hep-ph/0508081",
    archivePrefix = "arXiv",
    reportNumber = "CAFPE-63-05, UG-FT-193-05",
    doi = "10.1103/PhysRevLett.95.172502",
    journal = "Phys. Rev. Lett.",
    volume = "95",
    pages = "172502",
    year = "2005"
}

@article{Oller:2006jw,
    author = "Oller, Jose A.",
    title = "{On the strangeness -1 S-wave meson-baryon scattering}",
    eprint = "hep-ph/0603134",
    archivePrefix = "arXiv",
    doi = "10.1140/epja/i2006-10011-3",
    journal = "Eur. Phys. J. A",
    volume = "28",
    pages = "63--82",
    year = "2006"
}

@article{Hyodo:2008xr,
    author = "Hyodo, Tetsuo and Jido, Daisuke and Hosaka, Atsushi",
    title = "{Origin of the resonances in the chiral unitary approach}",
    eprint = "0803.2550",
    archivePrefix = "arXiv",
    primaryClass = "nucl-th",
    doi = "10.1103/PhysRevC.78.025203",
    journal = "Phys. Rev. C",
    volume = "78",
    pages = "025203",
    year = "2008"
}

@article{Ikeda:2012au,
    author = "Ikeda, Yoichi and Hyodo, Tetsuo and Weise, Wolfram",
    title = "{Chiral SU(3) theory of antikaon-nucleon interactions with improved threshold constraints}",
    eprint = "1201.6549",
    archivePrefix = "arXiv",
    primaryClass = "nucl-th",
    doi = "10.1016/j.nuclphysa.2012.01.029",
    journal = "Nucl. Phys. A",
    volume = "881",
    pages = "98--114",
    year = "2012"
}

@article{Mai:2012dt,
    author = "Mai, Maxim and Meissner, Ulf-G.",
    title = "{New insights into antikaon-nucleon scattering and the structure of the Lambda(1405)}",
    eprint = "1202.2030",
    archivePrefix = "arXiv",
    primaryClass = "nucl-th",
    doi = "10.1016/j.nuclphysa.2013.01.032",
    journal = "Nucl. Phys. A",
    volume = "900",
    pages = "51 -- 64",
    year = "2013"
}

@article{Guo:2012vv,
    author = "Guo, Zhi-Hui and Oller, J. A.",
    title = "{Meson-baryon reactions with strangeness -1 within a chiral framework}",
    eprint = "1210.3485",
    archivePrefix = "arXiv",
    primaryClass = "hep-ph",
    doi = "10.1103/PhysRevC.87.035202",
    journal = "Phys. Rev. C",
    volume = "87",
    number = "3",
    pages = "035202",
    year = "2013"
}

@article{Ramos:2016odk,
    author = "Ramos, A. and Feijoo, A. and Magas, V. K.",
    title = "{The chiral S = {\ensuremath{-}}1 meson{\textendash}baryon interaction with new constraints on the NLO contributions}",
    eprint = "1605.03767",
    archivePrefix = "arXiv",
    primaryClass = "nucl-th",
    doi = "10.1016/j.nuclphysa.2016.05.006",
    journal = "Nucl. Phys. A",
    volume = "954",
    pages = "58--74",
    year = "2016"
}

@article{Kamiya:2016jqc,
    author = "Kamiya, Yuki and Miyahara, Kenta and Ohnishi, Shota and Ikeda, Yoichi and Hyodo, Tetsuo and Oset, Eulogio and Weise, Wolfram",
    title = "{Antikaon-nucleon interaction and $\Lambda$(1405) in chiral SU(3) dynamics}",
    eprint = "1602.08852",
    archivePrefix = "arXiv",
    primaryClass = "hep-ph",
    doi = "10.1016/j.nuclphysa.2016.04.013",
    journal = "Nucl. Phys. A",
    volume = "954",
    pages = "41--57",
    year = "2016"
}

@article{Cieply:2016jby,
    author = "Ciepl{\'y}, A. and Mai, M. and Mei{\ss}ner, Ulf-G. and Smejkal, J.",
    title = "{On the pole content of coupled channels chiral approaches used for the $\bar{K}N$ system}",
    eprint = "1603.02531",
    archivePrefix = "arXiv",
    primaryClass = "hep-ph",
    doi = "10.1016/j.nuclphysa.2016.04.031",
    journal = "Nucl. Phys. A",
    volume = "954",
    pages = "17--40",
    year = "2016"
}

@article{Sadasivan:2018jig,
    author = {Sadasivan, D. and Mai, M. and D{\"o}ring, M.},
    title = "{S- and p-wave structure of $S=-1$ meson-baryon scattering in the resonance region}",
    eprint = "1805.04534",
    archivePrefix = "arXiv",
    primaryClass = "nucl-th",
    reportNumber = "JLAB-THY-18-2699",
    doi = "10.1016/j.physletb.2018.12.035",
    journal = "Phys. Lett. B",
    volume = "789",
    pages = "329--335",
    year = "2019"
}

@article{Lu:2018zof,
    author = "Lu, Jung-Xu and Geng, Li-Sheng and Ren, Xiu-Lei and Du, Meng-Lin",
    title = "{Meson-baryon scattering up to the next-to-next-to-leading order in covariant baryon chiral perturbation theory}",
    eprint = "1812.03799",
    archivePrefix = "arXiv",
    primaryClass = "nucl-th",
    doi = "10.1103/PhysRevD.99.054024",
    journal = "Phys. Rev. D",
    volume = "99",
    number = "5",
    pages = "054024",
    year = "2019"
}

@article{Oller:2019opk,
    author = "Oller, J. A.",
    title = "{Coupled-channel approach in hadron{\textendash}hadron scattering}",
    eprint = "1909.00370",
    archivePrefix = "arXiv",
    primaryClass = "hep-ph",
    doi = "10.1016/j.ppnp.2019.103728",
    journal = "Prog. Part. Nucl. Phys.",
    volume = "110",
    pages = "103728",
    year = "2020"
}

@article{Feijoo:2021zau,
    author = "Feijoo, Albert and Gaz da, Daniel and Magas, Volodymyr and Ramos, Angels",
    title = "{The K{\textasciimacron}N Interaction in Higher Partial Waves}",
    eprint = "2107.10560",
    archivePrefix = "arXiv",
    primaryClass = "hep-ph",
    doi = "10.3390/sym13081434",
    journal = "Symmetry",
    volume = "13",
    number = "8",
    pages = "1434",
    year = "2021"
}

@article{Sadasivan:2022srs,
    author = {Sadasivan, Daniel and Mai, Maxim and D{\"o}ring, Michael and Mei{\ss}ner, Ulf-G. and Amorim, Felipe and Klucik, John Paul and Lu, Jun-Xu and Geng, Li-Sheng},
    title = "{New insights into the pole parameters of the $\Lambda(1380)$, the $\Lambda(1405)$ and the $\Sigma(1385)$}",
    eprint = "2212.10415",
    archivePrefix = "arXiv",
    primaryClass = "nucl-th",
    doi = "10.3389/fphy.2023.1139236",
    journal = "Front. Phys.",
    volume = "11",
    pages = "1139236",
    year = "2023"
}

@article{Lu:2024ajt,
    author = "Lu, Yu and Jing, Hao-Jie and Wu, Jia-Jun",
    title = "{Phase Conventions in Hadron Physics from the Perspective of the Quark Model}",
    eprint = "2407.17131",
    archivePrefix = "arXiv",
    primaryClass = "hep-ph",
    doi = "10.3390/sym16081061",
    journal = "Symmetry",
    volume = "16",
    number = "8",
    pages = "1061",
    year = "2024"
}

@article{Hu:2017wli,
    author = {Hu, B. and Molina, R. and D{\"o}ring, M. and Mai, M. and Alexandru, A.},
    title = "{Chiral extrapolations of the $\boldsymbol{\rho(770)}$ meson in $\mathbf{N_f=2+1}$ lattice QCD simulations}",
    eprint = "1704.06248",
    archivePrefix = "arXiv",
    primaryClass = "hep-lat",
    reportNumber = "JLAB-THY-17-2445",
    doi = "10.1103/PhysRevD.96.034520",
    journal = "Phys. Rev. D",
    volume = "96",
    number = "3",
    pages = "034520",
    year = "2017"
}

@article{Mai:2019pqr,
    author = {Mai, Maxim and Culver, Chris and Alexandru, Andrei and D{\"o}ring, Michael and Lee, Frank X.},
    title = "{Cross-channel study of pion scattering from lattice QCD}",
    eprint = "1908.01847",
    archivePrefix = "arXiv",
    primaryClass = "hep-lat",
    reportNumber = "JLAB-THY-19-3017",
    doi = "10.1103/PhysRevD.100.114514",
    journal = "Phys. Rev. D",
    volume = "100",
    number = "11",
    pages = "114514",
    year = "2019"
}

@article{Hudspith:2024kzk,
    author = "Hudspith, Renwick J. and Lutz, Matthias F. M. and Mohler, Daniel",
    title = "{Precise Omega baryons from lattice QCD}",
    eprint = "2404.02769",
    archivePrefix = "arXiv",
    primaryClass = "hep-lat",
    month = "4",
    year = "2024"
}

@article{Mai:2017vot,
    author = "Mai, M. and Hu, B. and Doring, M. and Pilloni, A. and Szczepaniak, A.",
    title = "{Three-body Unitarity with Isobars Revisited}",
    eprint = "1706.06118",
    archivePrefix = "arXiv",
    primaryClass = "nucl-th",
    reportNumber = "JLAB-THY-17-2496",
    doi = "10.1140/epja/i2017-12368-4",
    journal = "Eur. Phys. J. A",
    volume = "53",
    number = "9",
    pages = "177",
    year = "2017"
}

@article{Mai:2017bge,
    author = {Mai, M. and D{\"o}ring, M.},
    title = "{Three-body Unitarity in the Finite Volume}",
    eprint = "1709.08222",
    archivePrefix = "arXiv",
    primaryClass = "hep-lat",
    reportNumber = "JLAB-THY-17-2554",
    doi = "10.1140/epja/i2017-12440-1",
    journal = "Eur. Phys. J. A",
    volume = "53",
    number = "12",
    pages = "240",
    year = "2017"
}

@article{Doring:2025phq,
    author = {D{\"o}ring, Michael and Khemchandani, Kanchan P. and Mart{\'\i}nez Torres, Alberto},
    title = "{Revisiting the three-kaon interaction and its relation with K(1460)}",
    eprint = "2511.02543",
    archivePrefix = "arXiv",
    primaryClass = "nucl-th",
    reportNumber = "JLAB-THY-25-4598",
    doi = "10.1103/4zf5-17p9",
    journal = "Phys. Rev. D",
    volume = "113",
    number = "3",
    pages = "034032",
    year = "2026"
}

@article{Feng:2024wyg,
    author = {Feng, Yuchuan and Gil, Fernando and D{\"o}ring, Michael and Molina, Raquel and Mai, Maxim and Shastry, Vanamali and Szczepaniak, Adam},
    title = "{A unitary coupled-channel three-body amplitude with pions and kaons}",
    eprint = "2407.08721",
    archivePrefix = "arXiv",
    primaryClass = "nucl-th",
    reportNumber = "JLAB-THY-24-4107",
    doi = "10.1103/PhysRevD.110.094002",
    journal = "Phys. Rev. D",
    volume = "110",
    pages = "094002",
    year = "2024"
}

@article{Yan:2025mdm,
    author = {Yan, Haobo and Mai, Maxim and Garofalo, Marco and Feng, Yuchuan and D{\"o}ring, Michael and Liu, Chuan and Liu, Liuming and Mei{\ss}ner, Ulf-G. and Urbach, Carsten},
    title = "{Emergence of the {\ensuremath{\pi}}(1300) Resonance from Lattice QCD}",
    eprint = "2510.09476",
    archivePrefix = "arXiv",
    primaryClass = "hep-lat",
    doi = "10.1103/vfr3-5lsb",
    journal = "Phys. Rev. Lett.",
    volume = "136",
    number = "14",
    pages = "141901",
    year = "2026"
}

@article{Feng:2026ixm,
    author = {Feng, Yuchuan and Culver, Chris and D{\"o}ring, Michael and Mai, Maxim and Alexandru, Andrei and Lee, Frank X.},
    title = "{Coupled-channel approach to isotensor pipipi scattering from lattice QCD}",
    eprint = "2601.16916",
    archivePrefix = "arXiv",
    primaryClass = "hep-lat",
    reportNumber = "JLAB-THY-26-4594",
    month = "1",
    year = "2026"
}

@misc{hadron,
        author = {Kostrzewa, Bartosz and Ostmeyer, Johann and Ueding, Martin and Urbach, Carsten},
        url = "https://github.com/HISKP-LQCD/hadron",
        howpublished = "https://github.com/HISKP-LQCD/hadron",
        title = {hadron: package to extract hadronic quantities},
        note = {{R} package version 3.0.1},
        year = "2020"
}

@article{PDG,
  author       = {Navas, S. and others},
  collaboration = {Particle Data Group},
  title        = {Review of particle physics},
  journal      = {Phys. Rev. D},
  volume       = {110}, 
  number       = {3},
  pages        = {030001},
  year         = {2024},
  doi          = {10.1103/PhysRevD.110.030001}
}

@manual{R-base,
  title        = {{R}: a language and environment for statistical computing},
  author       = {{R Core Team}},
  organization = {R Foundation for Statistical Computing},
  address      = {Vienna, Austria},
  year         = {2019} 
}
\bibliographystyle{unsrt}

\appendix

\onecolumngrid
\section{Spectrum octet irreducible representations}
\label{section:ap_gevp}

\begin{figure}[!h]
  \centering
  \includegraphics[width=0.6\linewidth]{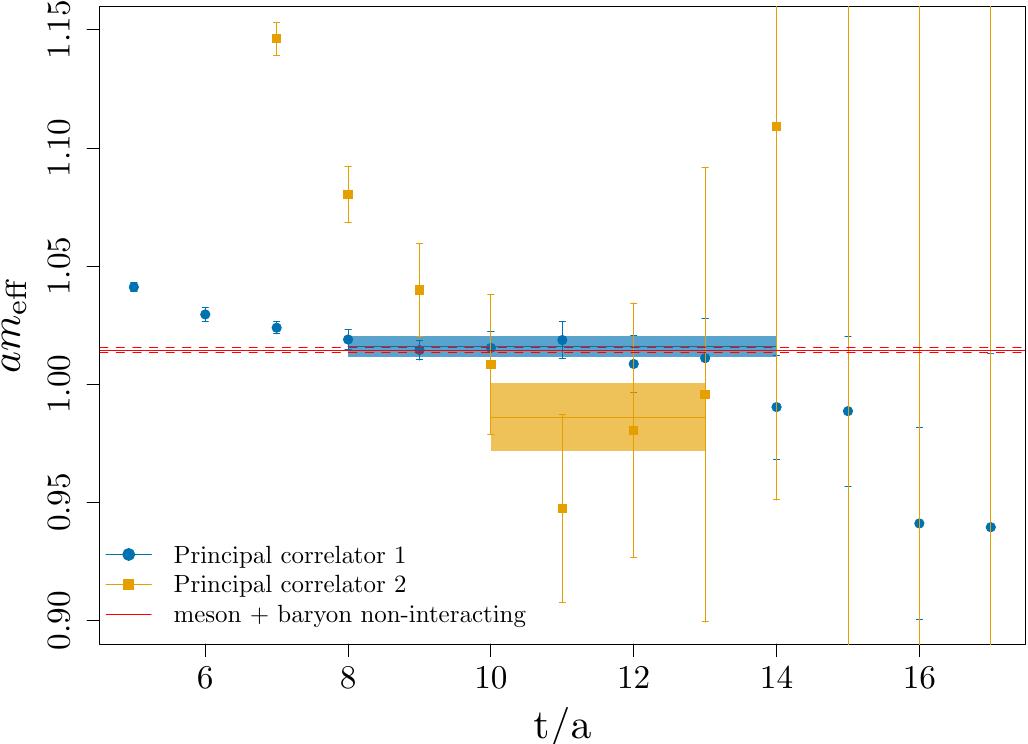}
  \caption{Effective mass of the principal correlators for the octet from the GEVP }
  \label{fig:gevp_octet}
\end{figure}

\begin{figure}[htpb]
  \centering
  \includegraphics[width=0.6\linewidth]{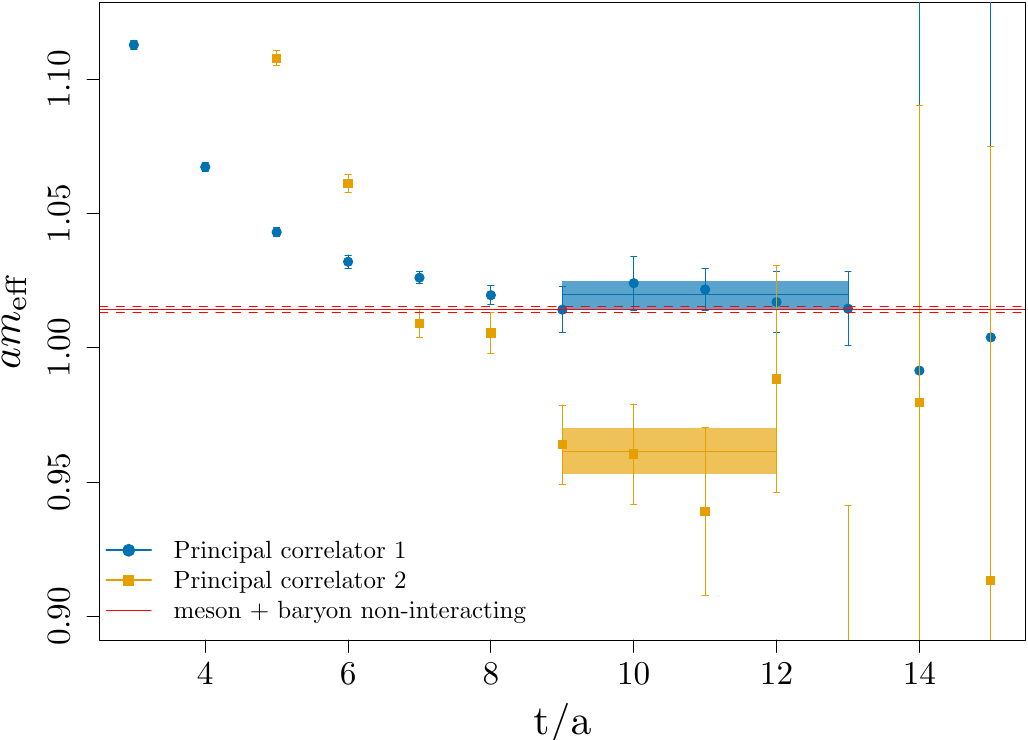}
  \caption{Effective mass of the principal correlators for the octet` from the GEVP }
  \label{fig:octet_prime}
\end{figure}

\clearpage
\section{Pole pictures}
\label{appendix:poles}

\begin{figure*}[h]
  \centering
  \includegraphics[width=0.99\linewidth]{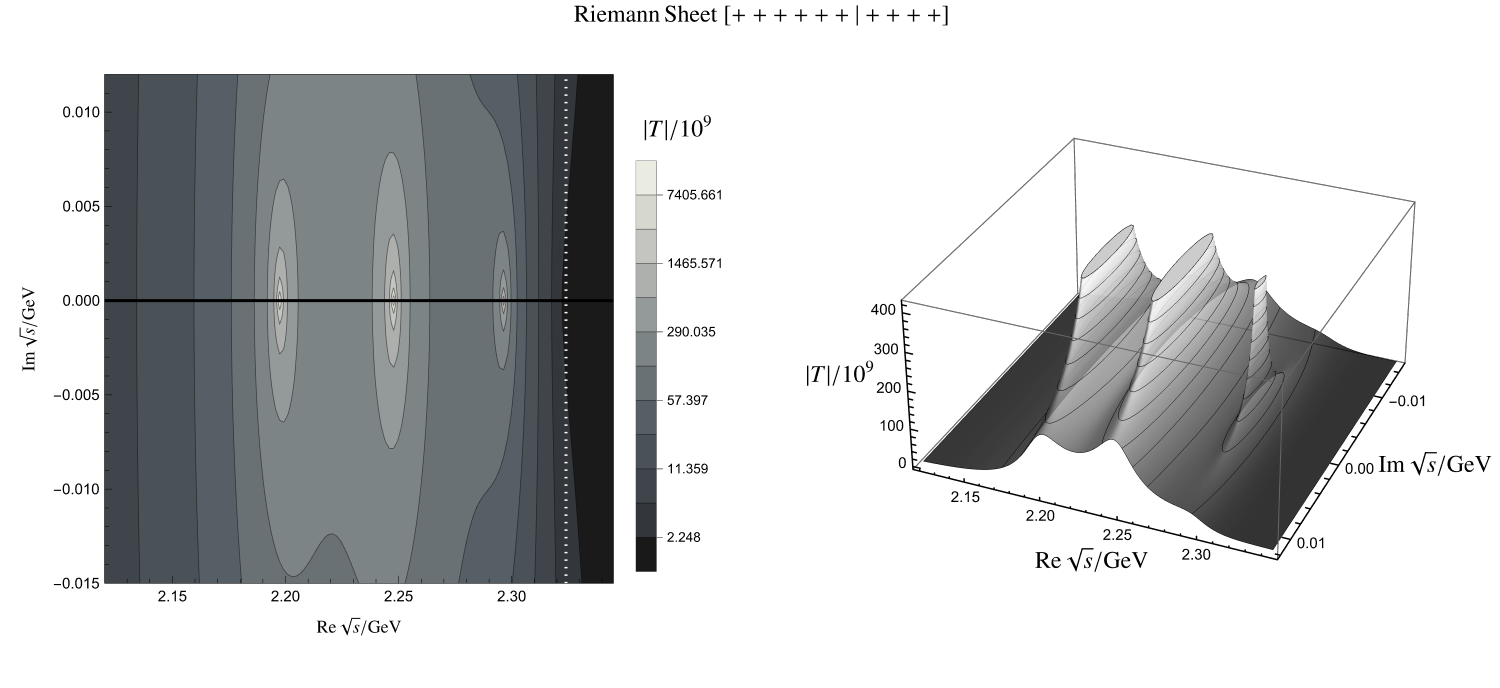}
  \caption{Fit $F_{16}$ at the $\mathrm{SU}(3)$ flavor symmetric point. Poles correspond to bound states}
\end{figure*}

\begin{figure*}[h]
  \centering
  \includegraphics[width=0.99\linewidth]{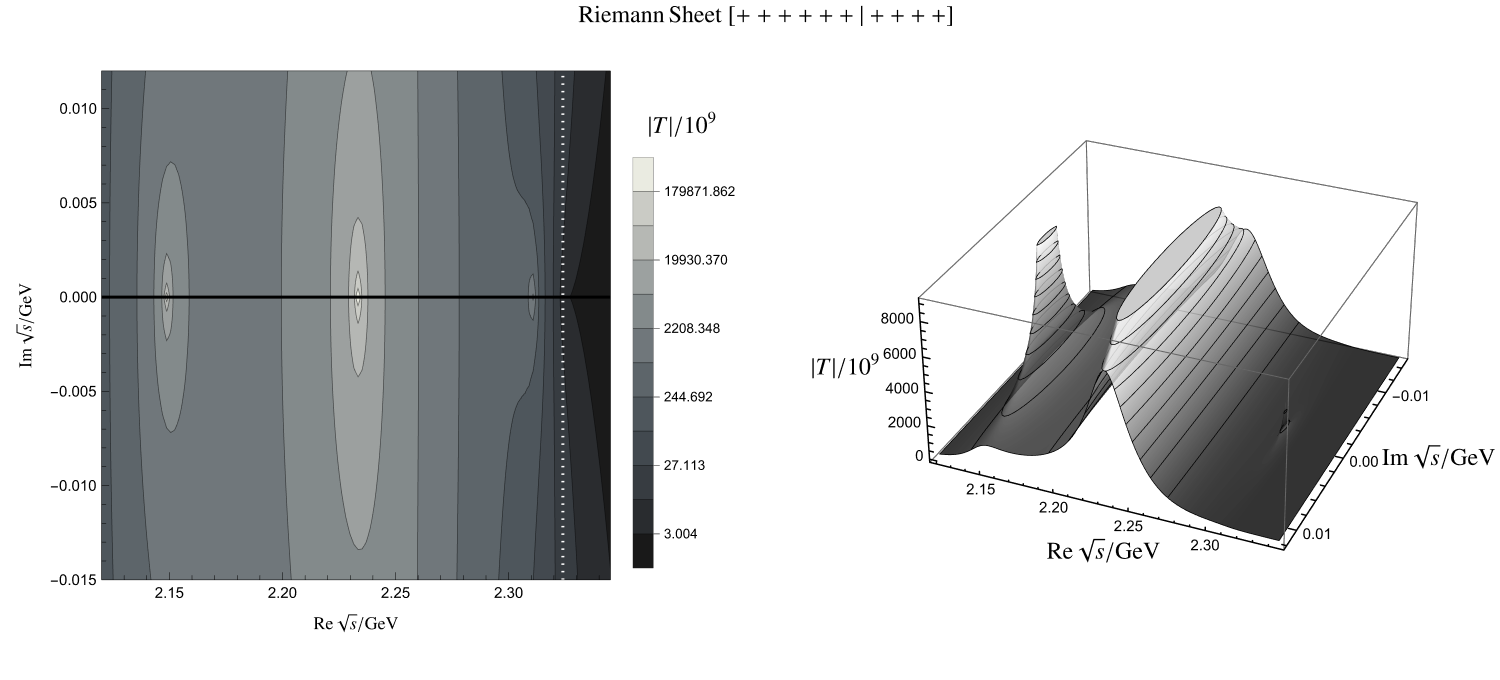}
  \caption{Fit $F_{16'}$ at the $\mathrm{SU}(3)$ flavor symmetric point. Poles correspond to bound states}
\end{figure*}

\end{document}